SWU for U and Me

Jeremy Bernstein

The intent of the sections that follow is to present the basic physics required to understand the workings of the gas centrifuge. The theory was first worked out during the Second World War and involved people like Dirac. I will provide references so that anyone who wants more detail will know where to find it. Gas centrifuges are at present the method of choice to separate the isotopes of uranium. That is a good motivation for studying the physics.

I.

The Iranian centrifuge program has brought up technical details about the separation of isotopes which have been heretofore the province of people from URENCO, and the like, who do this sort of thing for a living. Now, anyone interested in nuclear proliferation will run across the unit SWU- "separative work unit"- pronounced SWOO-which is a measure of how much an individual centrifuge can actually separate. One is therefore motivated to understand what this unit is. Here is a typical and rather good website

The equation defining Separative Work is:

$SWU = P \cdot V(N_p) + W \cdot V(N_w) - F \cdot V(N_f)$

Where P is the product amount, $N_p$ is the product concentration, W is the waste amount, $N_w$ is the waste concentration, F is the feed amount, $N_f$ is the feed concentration, and $V(x)$ is a value function that takes the form:

$V(x) = (2x-1)\ln(x/(1-x))$ where x is a given concentration

Where on earth does this bizarre unit come from? The purpose of this section is an attempt to explain.

The unit was first introduced by P.A.M. Dirac in 1941 in an unpublished paper on isotope separation. It is referred to in a seminal paper by

Klaus Fuchs and Rudolf Peierls written in 1942.[1] This work was no doubt turned over to the Russians bhy Fuchs who must have, as a consequence, realized that the British were interested in isotope separation. My discussion will follow closely this paper and that of the review article by Stanley Whitley.[2] These discussions begin with a statement of what the change of entropy is when a binary mixture of perfect gasses at a common temperature is separated into its components. Anything involving changes of entropy generated in the mixing and un-mixing of gasses seems also to generate much confusion, some of which I shared. The menace of the "Gibbs paradox" was always lurking in the background. So I will begin with a very elementary discussion of this.

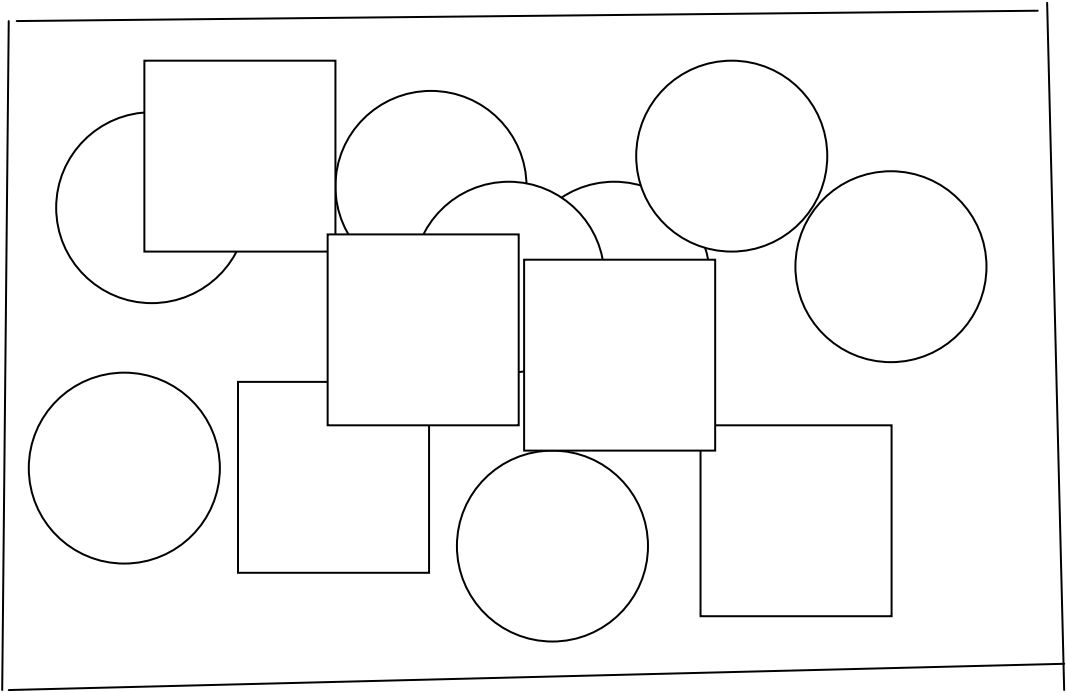

This amateurish diagram is meant to show a volume V filled with perfect gasses A and B in equal amounts and mixed together. The next diagram shows the same gasses separated by a partition in such a way that the volume is divided in half as are the number of particles of each type.

---

[1] K.Fuchs and R.Peierls Separation of Isotopes, Selected Scientific Papers of Sir Rudolf Peierls, World Scientific,Singapore, 1997 p. 303,
[2] S, Whitely, Rev.Mod.Phys, Vol.56, No1, January 1994.

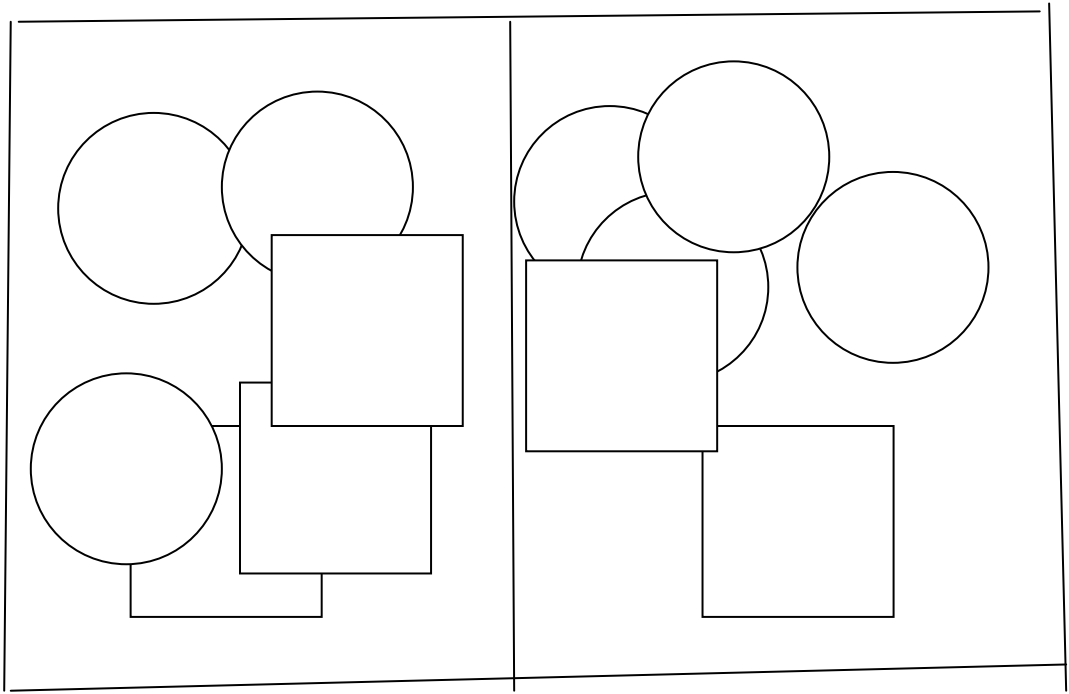

What is the entropy change?

The entropy associated with the first diagram is

S=kN/2ln(V/N/2)+kN/2ln(V/N/2) +C=kNln(2V/N)+C

Here N is the total number, V the total volume and C a constant that cancels out in the entropy change assuming the temperature is the same throughout. I will drop this constant. The entropy associated with the second diagram is for each box 2(N/4ln(V/2/N/4)) so adding the boxes we get kNln(2V/N). To no one's surprise there is no entropy change. Stirling's formula has been applied throughout. There are some very interesting additions to these formulae if you don't.

In the diagram below I replace the partition by a semi-permeable membrane.

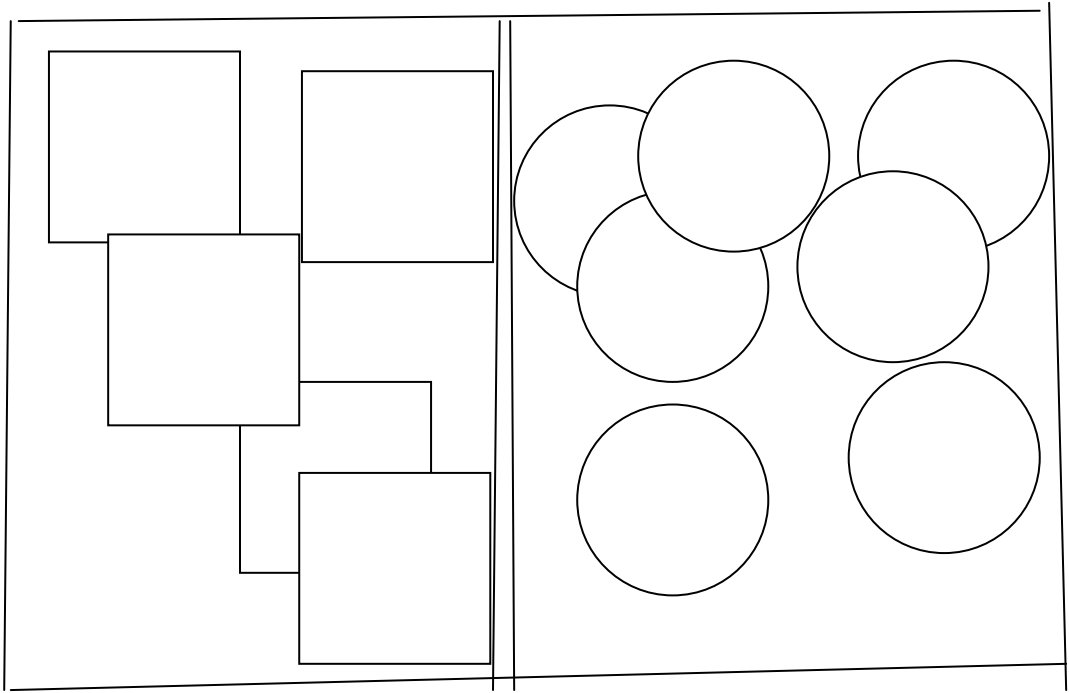

Now the entropy is $2kN/2\ln(V/2/N/2) = N\ln(V/N)$ so the entropy change is

$$Nk(\ln(2V/N) - \ln(v/N)) = k\ln(2^N).$$

I leave it to you Shannon fans to interpret this in terms of the information needed to do the sorting. The point is that the separation causes an entropy change. This explains the form of the entropy that Peierls and Fuchs write down. There is no log(2) term which in other contexts the Gibbs paradox requires. With $R=Nk$ and c the concentration of one of the components and hence 1-c the concentration of the other they write the difference between the entropy of the mixture as compared with the entropy of the constituents as

$$\Delta S = R(c\ln(c) + (1-c)\ln(1-c)).$$

In the operation of cascades of centrifuges each centrifuge produces a rather small amount of separation so that it is reasonable to introduce a small parameter d and expand $\Delta S$ for small d.[3] (Fuchs and Peierls consider a more general case to which I return in the next section). They suppose that the original sample is divided into two fractions which are not necessarily equal. Implicit here is that the division is into two equal parts.)

---

[3] I am grateful to Dick Garwin for pointing this out.

$$\Delta S \sim R\, d^2/2c(1-c),$$

To follow the next step as it is taken by these authors we need to do a small bit of combinatorics. Suppose we have a binary mixture which has a total of n particles with the two components having respectively $n_1$ and $n_2$ particles such that $n_1+n_2=n$, then how many pairs can be formed? The answer is

$$(n-1)n/2 = n_{like} + n_{unlike}$$

where the subscripts refer to pair that contain like and unlike particles respectively. Indeed

$$n_{like} = (n_1-1)n_1/2 + (n_2-1)n_2/2$$

while

$$n_{unlike} = n_1 n_2 = n_1(n-n_1)$$

Readers may entertain themselves by studying little pictures of samples and seeing that this works.

At this point Fuchs and Peierls do something that all the people who have followed them have repeated. They divide ΔS by c(1-c) and carry on. The only reference that I have found that gives a semblance of an explanation is indeed their paper. Indeed they have a footnote that reads

"The reason for this is that, with all usual methods, the work done by the device on the molecules {in separating them} is approximately independent of their nature. Of all possible pairs of molecules only the fraction c(1-c) are unlike ones, and only on those cases can the work done with the purpose of distinguishing them lead to any useful result. In all other cases it is wasted. Hence the factor c(1-c) in the efficiency."

This sounds good until one reflects that a centrifuge does not work on pairs but on individual particles. A demon that looked over the array and separated pairs this way might qualify for this remark, but I do not see its relevance for a centrifuge. I cannot explain what I do not understand. I can only tell you that the value function has the form that it does because this division is

made. Thus the entropy change per unlike pair, ΔY is given for small changes in concentration by

$$\Delta Y = R d^2/(c(1-c))^2.$$

The next and final step that leads to the definition of the value function is to ask for a function V(c) whose second derivative is $1/(c(1-c))^2$. Finding V(c) is trivial especially if you have a good integration program. Dropping the two additive constants on the grounds that you are only interested in differences then you have

$$V(c) = (2c-1)\ln(c/1-c)$$

This is the famous Dirac value function. I will in subsequent sections explain how this is applied to centrifuges.

## II

In section 'I', I attempted to explain how the value function of Dirac had been derived by Klaus Fuchs and Rudolf Peierls in their seminal 1942 paper on isotope separation. I will begin with a further discussion of this before going on in the next section to its application to the centrifuge. Consider the expression for the change in entropy produced by such a separation of two constituents in a binary mixture. As in ' I', I denote the concentration of one of the constituents by "c". Then the other has a concentration of 1-c. Then the entropy of the mixture, ΔS is given by

$$\Delta S = -R(c\ln(c) + (1-c)\ln(1-c))$$

I noted that if you introduced a semi-permeable membrane into the original volume and, if there was a fifty-fifty admixture of the two components-c=1/2- then, after complete separation, the total entropy of the separated components would be

$$S = k\ln(2^N)$$

where N is the total number of molecules. I remarked that Shannon buffs can interpret this in terms of the information needed to make the separation by a demon. I then returned to the original expression for the entropy and supposed a small change in the concentration ,'d' and expanded in a Taylor series.  This would produce a small change in the entropy, δS, given by

$$\delta S \sim Rd^2/2c(1-c).$$

This calculation has assumed that the original sample has been divided into two equal parts. I will consider the more general case at the end of the section. Fuchs and Peierls consider this change per unlike pair of molecules, the number of which is given by c(1-c). They then find a function V(c) whose second derivative is this change;ie, they solve the differential equation

$$d^2/dc^2 y(c)=1/(2c^2(1-c)^2)$$

This function turns out to be,

$$y(c)=(2c-1)\log(c/(1-c))+ac+b$$

where a and b the integration constants. They, following Dirac, fix these constants by insisting that if $c_0$ is the concentration of one of the components of the natural mixture then both y and its derivative must vanish at $c=c_0$. This gives them a form of the function

$$V=(2c-1)Ln((c/1-c)(1-c_c)/c_0))+(c-c_0)(1-2c_0)/c_0(1-c_0).$$

This differs from anything that you will find in  most treatments. An exception is the important book by Karl Cohen[4].In view of what I am about to say about the free energy, it seems  more reasonable to demand that y, and its derivative, vanish at c=1/2. This gives a=b=0 and  the common form of the value function, which Cohen, who considers this possibility, calls the "elementary value function," V(c) with

$$V(c)=(2c-1)\log(c/(1-c)).$$

---

This is what everyone uses without making any sensible comments about either where it comes from or where the integration constants have gone. Whitely's observation,

"Provided one is interested in only changes of this function, the arbitrary constants of integration are unimportant"[5] is typical.

Certainly no one can stop you-or Dirac-from defining such a function but what has it got with separative work-the work required to separate isotopes in a binary mixture?  To me, at least, their derivation on this point is totally opaque. The light began to dawn when I came across a brief 1978 article written by John B.Opfell entitled Free Energy Explanation of Separative Work.[6] Opfell is a chemical engineer and his arguments,  use concepts that, one gathers, are familiar to such people.  Opfell begins with an identity

$$(2c-1)\ln(c/(1-c)) = c\ln(c/(1-c)) - (1-c)\ln(c/(1-c)).$$

He then proceeds to show that this represents the differences of Gibbs free energy for the separation of the two components. Below is a plot of this function

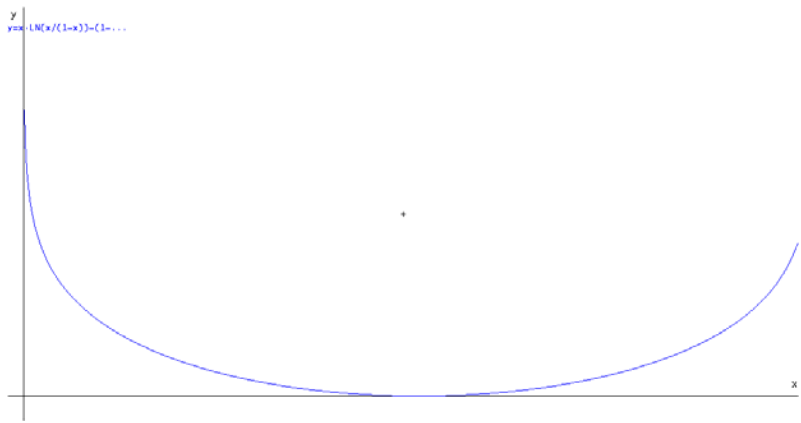

The net free energy difference is always positive, but the individual free energies may change sign as you either add to or subtract from a given component of the

---

[5] Whitely op cit. p.61.
[6] AiChE Journal (Vol.24, No.4) p.726.

mixture. Below is a plot of cln(c/(1-c)) which shows the sign change as you pass through c=1/2

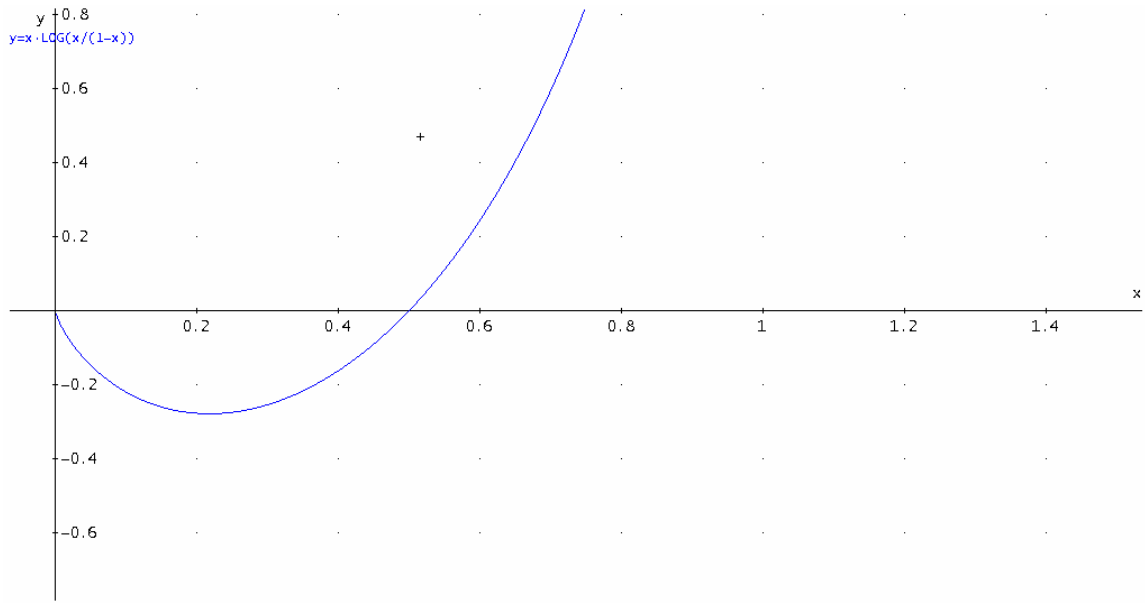

The point c=1/2 is very interesting. We saw above that the entropy is not zero, but we now see that the Gibbs free energy is zero at this point. In this irreversible process no useful work is needed. In a sense this is obvious because all we have done in our thought experiment is to raise and lower a membrane. This example was already studied by Gibbs in the 19th century and he came to the same conclusion.[7] There is an important lesson here. This thermodynamic exercise implies that we can separate a fifty -fifty mixture with no change in the free energy. The work is all supplied by the TdS. This drastically underestimates the amount of work you would have to do with a real centrifuge, although these use only about forty watts. This calculation ignores everything having to do with the real world-friction, heat loss etc, etc. In commenting about this, Dick Garwin noted that it is analogous to the Carnot Cycle description of the steam engine.

---

[7] I strongly recommend a wonderful article by E.T.Jaynes called The Gibbs Paradox. If you Google Gibbs +Jaynes you will find sites that have it. See for example http://bayes.wustl.edu/etj/articles/gibbs.paradox.pdf.

Marvin Minsky once showed me a drawing of a bird that he had made. It had a caption, " This looks like a bird, but no bird looks like this." The value function tells you the **minimum** work you have to do to separate isotopes. The actual work is quite another story.

Speaking of Dirac, Helmut Rechenberg was kind enough to Xerox Dirac's paper " The Theory of the Separation of Isotopes by Statistical Methods,which is in his collective works.  The paper is only three pages because most of it is not there. There is a much longer appendix which is nowhere to be found. My own guess is that MI6-or some such British agency-has classified the appendix. This may seem absurd unless you have dealt with British classified documents. Starting in June of 1945, ten German scientists, among them Heisenberg, were held in considerable comfort in a manor house near Cambridge called Farm Hall. A few years ago I visited it with Michael Frayn and was surprised by how small it was. In any event during the six months they were there all their conversations were recorded. Some or all were transcribed. These transcriptions were kept classified for several decades. Reading them now, it is impossible to understand why. The Germans knew next to nothing about nuclear weapons. But I digress.

Let me briefly recapitulate Dirac's argument. In the part of the paper I have He does this for small changes in concentration. He also makes the assumption that, initially, the two concentrations are very close to each other, which is certainly not the case for the uranium isotopes.  But he notes that in the appendix he does this more generally. Clearly Fuchs and Peierls had read this paper, to which they refer. They use the same notation although their methodology is quite different. Dirac's methodology bears a family resemblance to that of Dr. Opfell. Dirac wants to calculate the separation energy required to take a mass m  from the mixture, where m has a concentration c. Dirac says that this energy must be proportional to m, which seems reasonable, so he writes it as mf(c). This f(c) is going to be the value function. He then writes down the separation energy for a binary mixture with concentrations $c_1$ and $c_2$ under the assumption that $c_2$ is the concentration of the heavier mass isotope, $m_2$, and $c_1$ and $m_1$ are the corresponding quantities for the lighter isotope. Moreover, it is assumed that the two masses are very close together and concentrations are very close to each other. With these assumptions he derives two expressions for this separation energy that he equates assuming that $\delta c_2 = \varepsilon/m_2$ while-or whilst- $\delta c_1 = -\varepsilon/m_2$. With $m_1 \sim m_2$ we have $\delta(c_1+c_2) \sim 0$. The first expression is

$$m_1 df(c_1)/dc_1 \delta c_1 + m_2 df(c_2)/dc_2 \delta c_2 = \varepsilon d^2 f(c_1)/dc_1^2 \, (c_2-c_1).$$

The second expression for this quantity is

$$\varepsilon(c_2-c_1)/c_1^2.$$

Equating the two we have the equation

$$d^2 f(c_1)/dc_1^2 = 1/c_1^2$$

which integrates-dropping the constants- to

$$f(c) = -\ln(c).$$

Our previous expression morphs into this when c is much less than 1.

    While what Dirac does is very interesting, the condition that the c's are both very small and approximately equal is much too restrictive. He may well drop these conditions in the unavailable appendix. The condition that the change in the concentrations is small, is common to all these derivations. There are two schools of thought. One says that each step in the centrifuge cascade involves small changes in the concentration so using this in the individual steps may be all right. The second school says, yes but there are 50,000 steps so the errors are going to pile up. When I read these discussions I am reminded of one of Viki Weiskopf's jokes. I can still hear him tell it. A married couple that is having difficulties goes to the rabbi in the shtetl. First the husband speaks. After he does the rabbi says, "you are right." Then the wife speaks after which the rabbi says, "you are right." The husband the objects, "how can we both be right?" to which the rabbi responds, "you are also right." I have looked at some treatments that do not make this assumption and they are pretty complicated.

    I want to end this part of the screed with an analysis you will find in Fuchs and Peierls. For a transition in which the internal energy is not changed and the temperature remains constant, the energy required, $\Delta U$, is given by

$$\Delta U = T\Delta S.$$

We imagine that we have achieved a certain separation characterized by the concentrations c and 1-c. This state is characterized by an entropy

$$S(c) = R(c\ln(c) + (1-c)\ln(1-c)).$$

We now follow Fuchs and Peierls[8] and suppose that the sample is divided into two fractions characterized by amounts $A_1$ and $A_2$ such that $A_1 + A_2 = 1$. They initially have the same concentrations of the light isotope 'c'. We now suppose that a small change in the concentration is made in each section, changes which we characterize by $c+\varepsilon_1$ and $c+\varepsilon_2$ respectively. We can now write the change in S, using a Taylor series, as

$$(A_1+A_2-1) S(c) + (A_1\varepsilon_1 + A_2\varepsilon_2)dS(c)/dc$$
$$+1/2(A_1\varepsilon_1^2 + A_2\varepsilon_2^2)d^2/dc^2 S(c),$$

where

$$d^2/dc^2 S(c) = R/c(1-c).$$

The first term vanishes because of the condition on the A's. To understand the second term we have to be clear how the changes in concentration come about. The molecules do not disappear. They are simply moved about. The concentrations of light molecules are proportional to their number. The total number does not change so we must have
$A_1(c+\varepsilon_1) + A_2(c+\varepsilon_2) = c$ so that $A_1\varepsilon_1 + A_2\varepsilon_2 = 0$. Thus,
to second order

$$\Delta S = (A_1\varepsilon_1^2 + A_2\varepsilon_2^2) 1/2 R/c(1-c).$$

We treated above the special case in which $\varepsilon_1 = -\varepsilon_2 = d$
$$\Delta S = d^2/2c(1-c)\ R.$$
Thus
$$\Delta U = TRd^2/2c(1-c).$$

---

[8] They actually consider mixtures that have more than two components, but for the case that interests us, uranium enrichment, two suffice..

What is generally done is to divide this by c(1-c) to get the energy required per unlike pair. Thus

$T\Delta S/\Delta U = Tc(1-c)$.

This is a maximum for c=1/2. If we begin with a mixture and insert a semi-permeable membrane and completely separate the isotopes there is, as we have argued, no free energy expended. There is however an entropy change given by Rln(2). Thus you would have to supply an amount of work given by $T\Delta S=RTln(2)$.

I will end this section here. In the next ones I will apply these ideas to the centrifuge which is, after all, the object of the exercise.

III.

I must begin this section with a confession, or at least an explanation. I began this exercise after it had become clear to me that events in Iran might be reaching a dangerous and irreversible point. The newspapers were filled with statements about the Iran's uranium enrichment program using centrifuges. Some of these accounts contained terms like "SWU's" and how many of same the Iranians could produce and what they could do with them. Like any red-blooded physicist I immediately Googled "SWU". The resulting definition was totally incomprehensible to me. I decided that I would learn what this was all about. From my teaching days, I always found that teaching was the best way of learning. There is nothing like having to give a lecture in 24 hours to concentrate the mind. As Dr. Johnson put it, "when a man knows he is to be hanged it concentrates the mind wonderfully." In any event, sections I and II represent my attempts to carry out this part of the program. Section III, this one, was going to be the grand finale with the work of the first two parts, which concerned the general theory of isotope separation, applied to the centrifuge, and, in particular to the Iranian centrifuges. It is on the latter shoal that the ship has gone aground. No one who knows anything can tell you about it. It is all classified and seems to have been so for many years. In his masterly review article, Whitley[9] informs that he can tell us nothing about developments after 1962, because they are all classified.

In addition to the usual security paranoia, there is I think something else at work here. Enriching uranium is a highly profitable business now. The British, Dutch and German consortium URENCO announced that on May 2$^{nd}$, 2007, they delivered their two hundred millionth SWU. But they are in competition. There are at least three other major companies enriching and selling uranium. The largest of these-and apparently the largest in the world-is the Russian company Diatom. In their several centrifuge facilities they produce a total of something like 25 million SWU per year. Some of this is for their own consumption, but the rest is sold throughout the world. They seem to do an export business of about a billion dollars a year. They also export some of their technology to places like China and apparently also to Iran. There is the American company USEC and the French company Eurodif. The latter is a charming company that was founded in 1973 jointly by France, Belgium, Spain and Sweden. In 1974, Sweden pulled out and the Iranians bought their share. The Shah lent them about a billion dollars. After the revolution, the Iranians pulled out but they still own stock in the company because they have shares in a company that owns stock. It is claimed that the Iranians do not have a right to either the technology or the output. The former is rather academic because the company only now is switching from using diffusion to separate isotopes to using centrifuges. So all these companies have proprietary secrets as well as security interests. So we can't learn much of anything. What I can tell you about is the provenance of all these centrifuges. Some of this I learned directly from one of the people who was most responsible-the late Austrian physicist-engineer Gernot Zippe. If you don't know the story, you will I think find it interesting.

Although I spoke to Zippe several times, I was not able to learn from him much about his personal history. He was born in November 1918 and died in May 2008. I do not know his birthplace. He died in Germany. He did get a degree in physics in Austria. By some sequence of events which I could not learn from him, he became, during the war, a flight instructor for the Luftwaffe. Until he was eighty, he was still piloting planes. Then under some equally mysterious circumstances he was captured by the Russians and sent to a rather grim prison camp. To understand what happened next we have to back up a bit.

---

[9] Op cit

While it is well-known that we had a scientific mission-ALSOS-that was sent to follow the troops closely as they moved into Europe in order to learn what the Germans had done about making a bomb. The lead scientist was Samuel Goudsmit, a Dutch-born American physicist. It was a rather small mission. They did capture ten important German physicists including Heisenberg and send them to England where they were interred in Farm Hall. Their conversations were bugged for six months. They had absolutely nothing to teach us about nuclear weapons. What is less well-known is that the Russians had a much larger mission with some forty scientists. They knew exactly what they wanted because they had learned about both ours and the German program by espionage. They knew where the German uranium was and took a bunch of it. They also knew who might help them make a bomb and where to find them. One of their targets was Manfred von Ardenne-a German inventor who had made a lot of money on various electronic inventions. He had a private laboratory on his estate outside of Berlin financed by the post office. All during the war they worked on things like the electromagnetic separation of uranium isotopes. The very bizarre theorist Fritz Houtermans, who was there, conceived the idea of using what we called plutonium as a fissile material. The Russians went directly to von Ardenne's laboratory and made off with it and him lock stock and barrel. Houtermans was long gone. The rest were taken to Sochi in Georgia on the Black Sea. They were joined by others including Gustaf Hertz of the Franck-Hertz experiment-a Nobelist. They were instructed to separate uranium isotopes. They divided into three groups-a diffusion group headed by Hertz, an electromagnetic separation group headed by Ardenne, and a centrifuge group headed by an air plane scientist, a physicist named Max Steenbeck. They were given some old Russian centrifuges to work with which they soon realized were very inefficient. Steenbeck had never worked on centrifuges but he developed some theoretical work that was similar to Dirac's.

What happened next is also a little murky. Ardenne seems to have been given lists of prisoners of war so he could see if any of them being held elsewhere might be useful. Zippe's name and qualifications appeared and he was transferred and began working with Steenbeck. He too had never worked on centrifuges. The soon realized that to do anything useful they would need centrifuges whose rotation  speeds were greater than the speed of sound in air. Zippe took over the experimental part of the work with Steenbeck doing the theory, with a group of some forty people. It took several years to do it, but by 1956, when Zippe was released, they had made a new type of centrifuge. Zippe told me that the best centrifuge they built was made  with of aluminum tubes. It had a diameter of 100 millimeters. Its peripheral speed was about 350 meters a second- a little more than the speed of sound in air. This meant that the frequency of rotation was about 67,000 per minute.In the late 1930,s Jesse Beams and his collaborators separated chlorine isotopes with gas centrifuges that had peripheral speeds of some  500  meters a second. They weren't very reliable. Zippe and Steenbeck had available to them the pre-war publications of Beams but they wanted to make long, thin centrifuges which everyone thought was impossible. They did it and the Russian centrifuges could run  reliably for months. When Zippe was allowed to leave the Soviet Union in 1956 he was not allowed to take out any plans, but he had them in his head. He thought that the Russians had abandoned the idea of using centrifuges for separating uranium isotopes and was surprised to learn in 1990 that they had industrialized them, ultimately making millions of them.  During the next few years,  after he returned from Russia, he acted as a consultant to the German and Dutch programs which set about to develop what are known as "Zippe centrifuges". He called them the "Russian centrifuge."

In 1970, the British, Germans and Dutch signed what is known as the "Treaty of Almelo" named after the Dutch town where it was signed. This is what created URENCO. Just about this time a Pakistani metallurgist named Abdul

Quadr Khan joined the company in Holland and proceeded to steal the plans for the centrifuge and bring them back to Pakistan. This is not the place to go into this, except to say that he trafficked the plans in many countries including Iran. It is not for nothing that the "P" in the P1 centrifuges that the Iranians are used at first stands for "Pakistan." He also trafficked them to North Korea and we may someday learn what the North Koreans made of this. Part of the paranoia connected with the sharing of centrifuge information is certainly connected with this, but I don't think all of it is.

As far as I know, the first people to suggest using a centrifuge for separating isotopes were the British physicists, F.W.Aston and F.A. Lindemann. Aston was awarded the Nobel Prize in 1922 for using his invention of the mass spectrograph to separate isotopes. They published their paper in 1919[10]. Lindemann was a rather remarkable fellow. He tested an idea for pulling airplanes out of spins by trying it out in an air plane. During the war he became Churchill's science advisor and was made Lord Cherwell. This inspired one of my favorite bits of British academic doggerel

> Lord Cherwell, when the war began,
> Was plain professor Lindemann.
> But now, midst ministerial cheers,
> He takes his place among the peers.
> The House of Christ with one accord
> Now greets its newly risen Lord.

It helps to know that Christ Church was the name of the Oxford college to which Lindemann belonged.

In the 19th century centrifuges had been used to separate liquids like milk and cream. But Lindemann and Aston wanted to use a gas centrifuge to separate isotopes. This idea was attached as a kind of codicil to their discussion as to how to separate isotopes of neon in the stratosphere using gravity. If there

---

[10] F.A. Lindemann and F.W.Aston, Philosophical Magazine, 37,1919, 523. There are some typos in this paper which proves that even then you had to proof read carefully.

is a single mass of atomic weight M then the pressure is related to the density at height h by the equation

$p = RT\rho/M$ , and

$dp = -g\rho dh$ so that $d\rho/\rho = -Mg/RT dh$ so that

$\rho = \rho_0 e^{-Mg/RT\Delta h}$

The idea is that there is a height h at which the mixing by convection stops and, at that height, the density is $\rho_0$. $\Delta h$ is the height above that height where the density is determined by the force of gravity. They then assumed that there were two isotopes-they focused on neon which has two stable isotopes. They called the initial densities $\rho_0$ and $K_0\rho_0$. The differential equation for $K/K_0$ has the solution

$K/K_0 = e^{-g\Delta h/RT(M_1 - M_2)}$.

They took an atomic weight difference of 2 and a $\Delta h$ of a 100,000 feet where a specially designed balloon would fill itself with the ambient air. A small, but in principle measurable, shift in the atomic weights would manifest itself. They proposed that someone do the experiment. I do not know if anyone took them up. Then they added a codicil on the centrifuge.

Those of you who remember Stanley Kubrick's film 2001, will recall one of the astronauts jogging around a centrifuge. As I can testify, having visited the set several times, this was a real centrifuge built at the cost of about 300,000 dollars by the Vickers Engineering Group. It had a thirty eight foot diameter and a maximum peripheral speed of about three miles an hour. Kubrick told me that he was thinking of selling rides to help recover the cost. He asked me not to put that in the profile I was writing of him.

Aston and Lindemann did not live long enough to see Kubrick's film, but in 1919 they grasped the analogy between the centrifugal force in the centrifuge and the action of gravitation when it came to separating isotopes. The acceleration g is replaced by $v^2/r$ where r is the distance from the axis of the centrifuge. We can also write this in terms of the frequency $\omega$ as $\omega^2 r$. If $\rho_1$ is the number density of one of the species then the total mass of this species being accelerated is $\rho_1 m_1$. Thus the force equation for this species is

$d\rho_1 = -\rho_1 m_1 \omega^2/RT r dr$

which we can solve to find

$$\rho(r) = \rho(0) \exp(-M_1 \omega^2 r^2 / 2RT)$$

where we are measuring from the axis. If we call K the ratio of the two densities then

$$K(r)/K(0) = \exp(-(M_1 - M_2) \omega^2 r^2 / 2RT).$$

It is instructive to evaluate the argument of the exponential for the parameters of the best Zippe centrifuge. The mass difference in gram moles between the uranium isotopes is about 3, the peripheral speed is $3.5 \times 10^4$ cm/sec, $R = 8.3 \times 10^7$ ergs/degree K. What to take for K? The material that is centrifuged is the gas uranium hexafluoride, a pretty corrosive compound. The advantage of using "hex" is that there is only one naturally occurring stable isotope so that the only isotope separation is with the uranium. Below is a thermodynamic curve for hex,

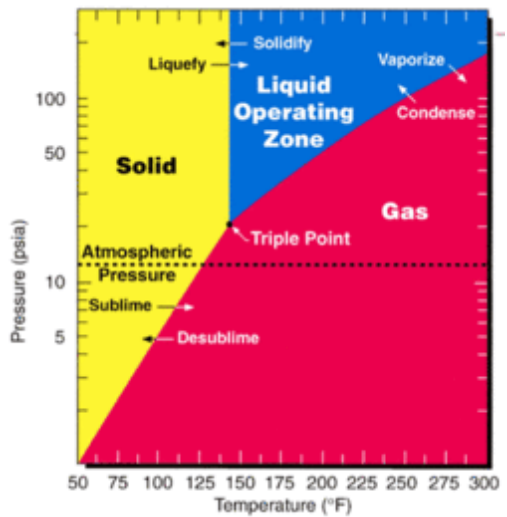

The triple point is at 337K so that is what we shall take although I suspect that the actual temperature is higher. If we call the parameter in the exponential 'd' we have

$$d = 3(3.5 \times 10^4)^2 / (8.1 \times 10^7 \times 337 \times 2) \sim .06$$

This would predict a small difference between the isotope ratio in the center and the edge of the centrifuge. Incidentally this is what Lindemann and Aston concluded for their hypothetical centrifuge. They note that "Separation by this method therefore seems possible though difficult and costly."[11] So it turned out to be.

In what follows I will take d as small. . We can now make us of the work of the previous sections. You will recall that the value function was written in terms of the concentrations c as

$$V(c)=(2c-1)\log(c/(1-c))=(c-(1-c))\log(c/(1-c))$$

If I call the argument of the exponentials for the two masses x, and y respectively then we have in this case

$$V=(e^x-e^y)/(e^x+e^y)\log(e^x/e^y).$$

Whether these exponentials are positive or negative depends on which isotopic mass you take to be heavier. If we let y=x+d and expand in d we find

$$V \sim d^2/2 = ((M_1-M_2)\,\omega^2\,r^2\,/2RT)^2/2$$
$$= ((M_1-M_2)\,v^2\,/2RT)^2/2$$

In the example we just look at this would give us

V~.002 How do we use this? From the previous discussion if we multiply V by RT we have the thermodynamic energy of separation of one mole. It is more interesting to ask for energy of separation of one kilogram. Each atom of U-235, for example, has a mass of $3.9 \times 10^{-25}$ kilograms and each mole contains $6 \times 10^{23}$ atoms. Putting in the numbers there are about four moles per kilogram. To get the thermodynamic energy of separation of a kilogram, we must multiply the above by the four moles and by RT which gives about 8 joules/degree Kelvin xT for each kilogram. If we put T=337 we get about 5 joules per kilogram. This is minuscule compared to the actual amount of energy it takes for a real centrifuge to separate a kilogram which is in the hundreds of mega joules. So the efficiency of these centrifuges is very small indeed. Most of this energy would be expended if a given centrifuge did not separate anything, for example if the gas had a single isotope. The energy is expended to keep the centrifuge turning. Nonetheless, the energy expenditure is orders of magnitude less than that required by other separation methods such as diffusion or electromagnetic separation. The electrical energy used at Oak Ridge in the electromagnetic separation of isotopes was greater than the energy produced by the Hiroshima bomb.

---

[11] Op cit p.532.

Well we started off by trying to understand the definition of the SWU-separative work unit. What a long strange trip it has been. We first needed to understand the value function and that took some doing. The SWU is defined in terms of the value function. The value function is dimensionless so the SWU, which is defined as LV(c) has whatever dimensions L has. In now many cases that I have seen does L have the dimensions of 'work.' In some cases L is in moles or kilograms or tons and in some cases it is in various rates. When you see an SWU quoted you have to pay attention to the small print. I am going to begin by considering a very symmetric situation. Below is a rendering that you will find on the web of what it is claimed to be the Zippe centrifuge.

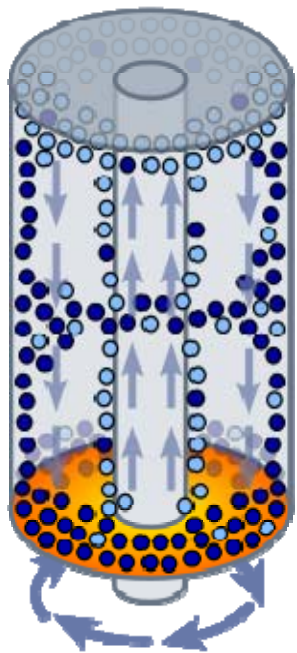

Actually it could be anybody's centrifuge. There are more detailed diagrams of what are said to be Zippe centrifuges on the web. I once asked Zippe of he had a simple diagram and he said he did not. Zippe and Steenbeck had the advantage of knowing nothing about any of this when they started. But they were very smart and highly motivated. They were told to separate isotopes or else. They did have some literature available to them-notably some articles of Jesse Beams, but they made important innovations and put everything together in a better way than anyone else had.

In the rendering above the light circles are U(235) and the heavy circles are U(238). The two are flowing in counter currents. The U(235) enriched current is captured by what are known as "scoops"-not shown, and the residual constitute the "tails." I am going to suppose that L moles have been divided equally between these two. Moreover I am going to assume that the enriched stream has a concentration given by c/a while the tail has a concentration of ac. The molar SWU which is the sum or two terms-one for the enriched stream and one for the tails can be written in this case as

$$SWU_{molar} = L/2(\,(a-1)/(a+1)Log(a)+((1/a-1)/(1/a+1)Log(1/a)=L(a-1)/(a+1))Log(a)$$

If a-1 is small then in this case
$$SWU_{molar} \sim L/2(a-1)^2$$
We have argued that the value function is proportional to the difference of the free energies. If we multiply the above by RT we come back to the expression for the thermodynamic energy needed to separate L moles into a division of L/2 moles with a small change in the concentrations.

In the more general case if we call P the product and T the tails in .say, moles with concentrations $c_P$ and $c_T$ the molar SWU is defined as

$$SWU_{molar}=PV(c_P) +TV(c_T).$$

Up to this point we have considered the situation in which a certain fixed amount is being separated. In reality what happens is that new material is being fed into the centrifuge all the time. This is above all true if the centrifuges are being operated in a cascade. If this feed just balanced the separation we wouldn't get anywhere, so this contribution to the SWU is negative. With F and $c_f$ referring to the feed we get the final statement of what a SWU is, namely

$$SWU_{molar} = PV(c_p) + TV(c_T) - FV(c_f)$$

I am going to stop this section here and in the next one I will begin to apply what we have learned to the real world.

IV.

Woody Allen was once asked for his views on the after life. "Can you get a good steak there?" he asked. In this section I will attempt to deal with the real world and the same question applies. First let me make it clear why the actual production of enriched uranium is such a difficult problem even if you assume that you have a good functioning individual centrifuge   To this end I am going to apply the work of the last section to compute how many SWU it would actually take to do something useful such as making a kilogram of highly enriched uranium.[12]

You begin with two equations that in one way or another express the conservation of matter. The notation is the following. Let 'F' be the feed into the centrifuge in kilograms, 'P' the product you want in same, and 'W' the waste also in kilograms. Then

$$F = P + W.$$

The second equation involves the molar fractions of the desired isotope in each of these parts, $x_f, x_p$ and $x_w$. The quantity $x_f$ we will take as .007, which is what you would get from a uranium mine. In any event the second equation is

$$x_f F = x_w W + x_p P$$

In what follows we shall set P=1kg. We can now solve
And write per kilogram

$$W/kg = (x_f - x_p)/(x_w - x_f)$$
$$F/kg = (x_w - x_p)/(x_w - x_f)$$

---

[12] I am very grateful to Jacob Bigeleisen for spelling out how this works in practice. Dick Garwin supplied a table in which the work is done for you.. But using it to find out how it works is a little like reverse engineering a three-star Michelin meal to see how the chef cooked it.

We need these quantities to compute the SWU in terms of the value functions. So using the work of the last section we have

SWU/kg=$V(x_p)+WV(x_w)-FV(x_f)$,

where I this case I have taken P=1 and as before

$V(x)=(2x-1)\ln(x/(1-x))$.

We can now put in some numbers and see what happens. I will take some values out of Garwin's Excel spread sheet because these give some idea of what the problem is. I will do my own arithmetic which will be less accurate than his. He considers the case in which $x_p$=0.95-highly enriched uranium indeed- $x_w$=.0025 and $x_f$=.007. (As I said ,Garwin takes more decimal points.) With these numbers I find per kilogram W=210 and F=211. The difference is the product which by construction is one kilogram. Next we must compute the value functions and take the weighted sum. With my arithmetic I find $V(x_p)$=2.65,$V(x_f)$=4.89, $V(x_W)$=6.0. The weighted sum is then

SWU/kg= 2.65+210x6.0-211x4.89/kg=231/kg.

Now we get to the point. The Iranians were using P1 centrifuges that they got at least conceptually and in parts from A.Q.Khan. There have been traces of HEU found on them which presumably reflects their provenance. Different values have been given for how many SWU per year these centrifuges produce. I have seen as low as 2 and as high as 5. Let us split the difference and take 3. I would then take 77 years for one of these centrifuges to produce enough SWU to produce 1 kilogram of HEU. Clearly this is not the way to go. Note that if we wanted to produce a kilogram of say 4 percent enriched uranium-LEU- the number drops to 5.7. It is also instructive to start with LEU and ask with the same centrifuge how many SWU are needed to make a kilogram. The number is about 73. There is clearly a moral here. Before I turn to what is really the heart of this screed, cascades I will tell you a story.

I am notoriously bad at arithmetic. It may be genetic. My scientific life was saved when Hewlett-Packard came out with their first scientific pocket calculator. I immediately rushed off to 47$^{nd}$ Street Camera-a famous store in those days run by Hasidim from Brooklyn. These people were much more interested in arguing with you than selling things, which is probably why they went out of business. I bought my calculator from a Hassid and the usual Talmudic discussion began, "Tell me," he asked, "is it true that with this calculator you can save a lot of time.?" "God yes," I answered. "Tell me," he asked, "what do you do with the time you save?"

To make a dent in the production problem you must run tens of thousands of centrifuges in some sort of cascade-a connected array of centrifuges. If you had, say, 50,000 centrifuges, and the production added up, you could make a kilogram in a jiffy.

The designing of cascades is both a science and an art. If you took some of the line drawings I have seen of cascades and signed them Mondrian you could probably sell them for a fortune. The photographs of actual operating cascades with their banks of elegant spinning centrifuges are quite beautiful. I make no pretense about the depth of my knowledge of this subject. But it is clear to me that there are some very,very smart people who have spent a career designing these things. I will try to give a bit of the essence. I hope the reader will come away sharing my feeling that this is a very elegant subject. My main objective, as Bohr unforgettably put it, will be not to speak more clearly than I think. I will deal only with what are known as "ideal" cascades-the meaning of the term will become clearer in the course of time, made up of centrifuges that divide the product and waste symmetrically. If $\alpha$ is the enrichment factor and R the concentration then the enriched stream is enriched by $\alpha R$ while the tails-waste- have a loss of $1/\alpha\ R$. Therefore, $\alpha$ is a number greater than one. As we proceed we are going to assume that $\alpha-1 \ll 1$. We are also going to assume mass conservation at every stage, something we have already used. There are no leaks. Suppose we call the abundance ratio of the two isotopes in the feed stream, R. Then 1-N, will be the molar ratio of the other and R=N/(1-N). It is this ratio that is going to change. If the feed stream is split in two then, say, the lighter isotope will have a concentration N' in one branch and N'' in the other such that

$$N' + N'' = N$$

Each branch will have its own R which is related to the corresponding N by the equation

$$N = R/(1+R).$$

But the branches will not divide equally. We suppose that the division is characterized by a fraction θ so that one branch changes by θ moles and the other by 1-θ moles. I chose θ, which is known as the "cut" as my symbol to correspond with the notation of Whitley in case anyone wants to refer to it. Moreover, we suppose, as noted, that concentrations are changed by αR and R/α. We can now write the mass conservation equation using the relation between N and R. So for N' we have

$$N' = αR/(1+αR).$$

For N'' we have

$$N'' = (R/α)/(1+R/α).$$

We can now use the conservation of mass

$$θN' + (1-θ)N'' = N$$

to yield the equation

$$θαR/(1+αR) + (1-θ)R/α /(1+R/α) = R/(1+R).$$

This ca be solved for θ; ie,

$$θ = (αR+1)/((α+1)(R+1))) = (N(α - 1) + 1)/(α+1)$$
$$\sim 1/(α+1)$$

$$1-θ = (α - N·(α - 1))/(α+1) \sim α/(α+1)$$

We note that for α~1, the parameter θ is independent of R. This will simplify the study of cascades in this limit. As we have defined things the waste stream is somewhat larger than the product stream, which is usually the case.

Note that if there is a fifty-fifty separation in the ideal case, then α=1 and all the centrifuge does is to split the stream in two without enrichment. Not worthwhile.

In the design of cascades there are two kinds of hookups-"parallel" and "serial"-which are combined in various ways depending on what outcome the designer wants to achieve. Below is a crude serial connection.

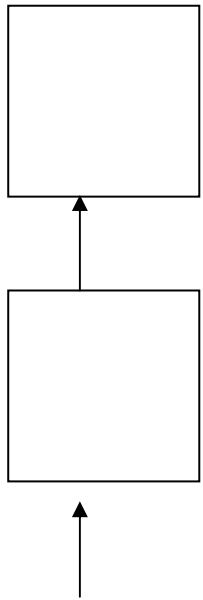

The essential point is that the output of one centrifuge is fed into another. If each centrifuge can produce nSWU then by definition 2 of them can produce 2nSWU. The SWU add up linearly. Does this mean that this configuration will produce two times the product of one centrifuge. The answer is clearly no. In the notation we have previously employed if L moles per unit time are fed into the bottom centrifuge then the enriched output will be θ(αR)L. The next stage has a net enrichment of α²R but the amount enriched will be θ( α²R)θ(αR)L. You can persuade yourself that the output will steadily decrease as more and more centrifuges are added to the string and the concentration of the desired isotope goes up. The result is certainly not additive.

The parallel connection is one in which the product of one centrifuge is not fed into another.

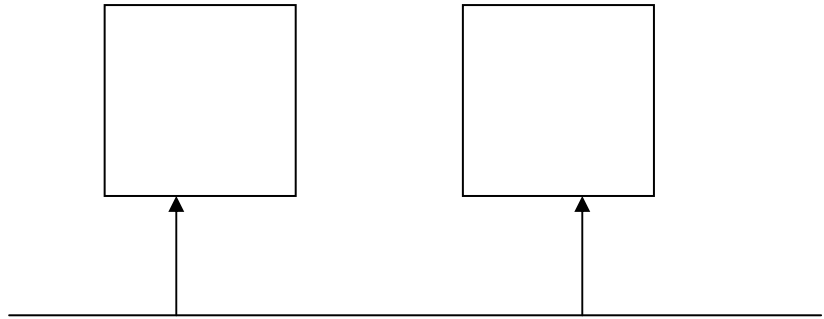

Here the outputs are additive. This raises the question in connection with the Iranian centrifuges or indeed any set of centrifuges but in the case of Iran it has some urgency. We know roughly how many SWU per year each centrifuge can produce. We know how many SWU is takes to produce one kilogram of highly enriched uranium- HEU. If we are told only how many centrifuges are in a cascade can we say how much HEU is produced. The answer is clearly no unless we know how the cascade is being designed. As I will now try to explain, the assumption implicit in these numbers is that the Iranians have "ideal" cascades. Any other configuration would produce fewer SWU, so what these numbers indicate is the *possible* enrichment and not the actual enrichment which is certainly less.

In the last ssection I presented the common figure which is used to illustrate what is claimed to be the Zippe centrifuge. I give it again below

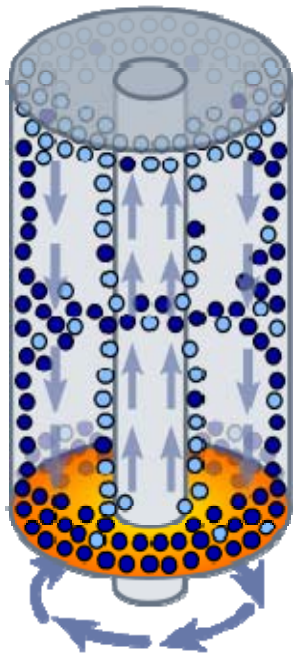

I said noted this diagram depicts a centrifuge-sort of-but it certainly does not depict what is specifically a Zippe centrifuge. The diagram shows a centrifuge that is short and squat. The Zippe centrifuge is long and thin. Their diameters are a few centimeters while their heights are something like a half a meter.This diagram shows no specific suspension-how it is connected at the bottom-on what does it spin-and how is it connected at the top. Zippe told me that the "Russian" group had available to them some of the reports by Jesse Beams with whom Zippe would later work. I am not sure which reports they had but it is plausible that they would have included Beams's 1938 review article-"High Speed

Centrifuging"[13] In this article Beams not only gives a history of the development of the modern centrifuge but discusses magnetic suspension of the top, which had been developed the previous year by F.T.Holmes, also of the University of Virginia. This enables an almost frictionless suspension of the upper centrifuge. The bottom is set on a ball bearing or a needle. The diagram shows what is known as a counter current centrifuge. The waste which has had some of its U-235 content removed flows in a downward current, while the product, which is now enriched flows upward. These currents can be extracted by "scoops." The idea of using countercurrents in centrifuges was introduced by Harold Urey in the late 1930s. He took over an idea that had long been applied to isotope separation by distillation. Whitley does point out important modifications that were made by the Zippe group but these were in the way of improving something that was already in use. But if you put enough of these improvements together you effectively create something quite new.

        The diagram below which was kindly supplied by Carey Sublette and seems to have been taken from a report by Zippe and Beams gives a better sense of what a Zippe centrifuge is.

---

[13] Reviews of Modern Physics, **10**,1938, 245.

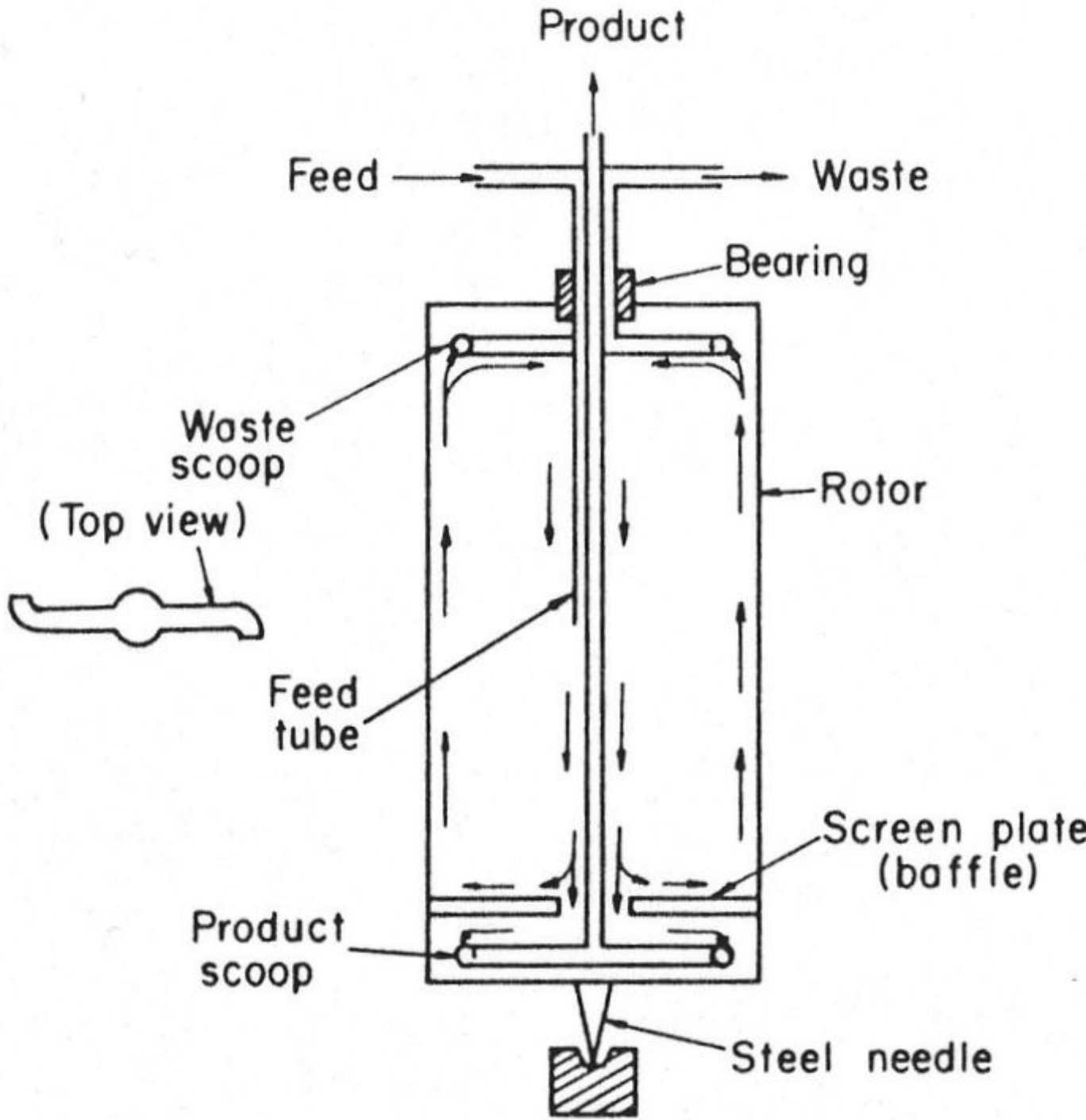

Before I turn to a discussion of SWU for a simple cascade model, I want to make a few more general observations. First, what do SWU do for you? I would put it this way. Suppose that you have a "task" that you want a single centrifuge or a cascade to do for you, how much separative work or power is required to do that task? That is a very different question from asking how many SWU per year some cascade can produce? What in this context is a "task"? A task is given some quantity of feed with some specified concentration of U-235, then separate this incoming stream into-say two-outgoing streams-product and waste-with specified concentrations of U-235. That is the task and to it there is a required SWU. Let me give three examples which I have used Garwin's wonderful ExCEL spread sheet to compute. In all these tasks the x with the subscript is the concentration of U-235 product, feed and waste

Task 1
$X_P=0.95$
$X_f=0.007$
$X_w=0.0025$

SWU/kilogram of product=221.
This task was producing highly enriched uranium-HEU-from hex with the natural concentration of U-235.

Task 2
$X_p=0.04$
$X_f=0.007$
$X_w=0.0025$

SWU/kilogram product=6.
This task was to produce lightly enriched uranium-LEU-from natural hex.

The final task-and perhaps the most interesting in terms of proliferation is to start with LEU and produce HEU.

$X_p = 0.95$

$X_f = 0.04$

$X_w = 0.0025$

SWU/kilogram product = 73.4

This reflects the often stated point that it is much easier to start from LEU and go to HEU then it is to start from scratch .

Why is this? I think of it in terms of my separating demon which looks for unlike isotopic pairs of uranium molecules to separate. The number of these goes as $x_f(1-x_f)$. This function starts at zero when there is no concentration of uranium-235 in the feed, as then rises to a maximum when there is a fifty-fifty mixture. The more unlike pairs there are, the easier is the separation and the smaller the SWU. If you put in the numbers I have been using you will see this qualitative behavior reproduced, although the quantitative answers are somewhat different.

I will now give a simplified discussion of SWU for cascades. The simplified discussion is hard enough-at least for me. I leave the more general discussion to the experts. Below is one centrifuge.

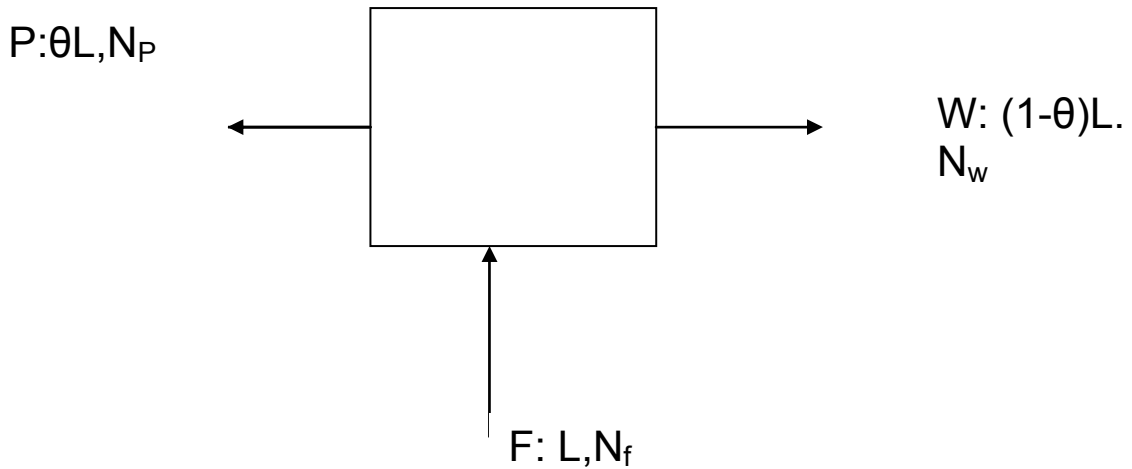

From our previous definitions the SWU, U, is given in terms of the feed L and θ by

$$U = L(\theta V(N_p) + (1-\theta)(V(N_w) - V(N_f)))$$

where as usual

$$V(N) = (2N-1)\ln(N/1-N)).$$

The dimension of this SWU is the dimension of L. If the feed is in kilograms the so is a SWU. Frequently reference is made to "separative power." This is also computed from the formula above when L is given in mass/per unit time. In the cases that we are interested in the usual unit is kilogram/year. We want to derive a formula for the change in U when N is changed by small amounts. This will tell us how much separative work is needed to make such a change. I am going to follow the treatment given in Karl Cohen's 1951 book with the long title that begins "The Theory of Isotope Separation…" The book is probably the defining monograph on this subject, I was put off of it at first because he does not take any prisoners-you can hit the wall easily unless you have familiarized yourself with some of the basics. The book originally published by McGraw Hill has long been out of print. In some desperation I located a place where they will make you your own copy for $61. It is called UMI-Books on Demand and their phone number is 734-761-4700. I do not know anything about them but they do excellent work. In deference to them and Cohen I will use his notation for the quantity we are looking for δU. What Cohen does is to expand δU in a Taylor series around the N for the feed. This is the .007 for natural uranium if this is where you start from. In a cascade the feed composition varies from stage to stage The idea is that in any stage the departures are small. Thus

$$\delta U = V(N)[L(\theta + (1-\theta) - 1) + dV(N)/dN\{\theta L(N_p - N) + (1-\theta)L(N_w - N) + 1/2\, Ld^2 V(N)/dN^2\{\theta(N_p - N)^2 + (1-\theta)(N_w - N)^2\}\}]$$

The coefficient of the first term vanishes byaddition, and the coefficient of the first derivative vanishes because of the conservation of mass. Here is where we use that condition explicitly. The second derivative is an old friend. See Section I. Integrating it is what got us the form of the value function in the first place. Thus

$$d^2 V(N)/dN^2 = 1/(N(1-N))^2$$

To deal with the rest of the expression, we are going to need to express $N_p$ and $N_w$ in terms of N. Here again we shall use the "ideality" of this example. Thus

$$N_p - N = \alpha R/(1+\alpha R) - R/(1+R))$$

By re-arranging this can be written as

$$N_p - N = (\alpha - 1)(N(1-N_p)) \sim (\alpha - 1)(N(1-N))$$

The last expression uses the assumption that each stage makes only small changes in the concentrations. The quantity α-1 is usually called in this literature, ε. Finally we can use the mass conservation equation once again

$$\theta(N_p - N) = -(1-\theta)(N_w - N)$$

to arrive at one of the most famous equations in this subject; ie, the amount of separative work required to make this change in concentration, δU, is given by when α~1

$$\delta U = L\theta/(1-\theta) \, 1/2\varepsilon^2 = L/2(1-\alpha)^2/\alpha$$

I am going to apply this to the toy ideal cascade that you will find in Whitley. This is the diagram below-the Mondrian.

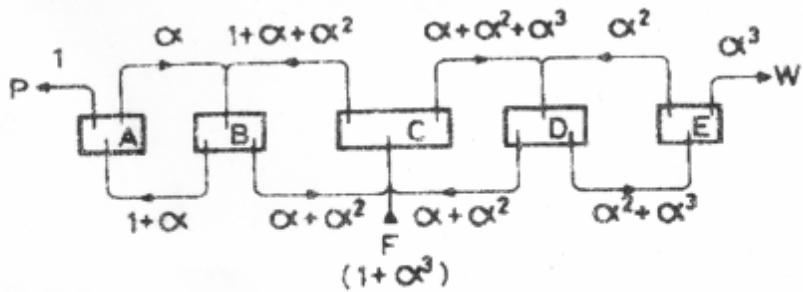

FIG. 16. Simple symmetrical cascade.

To interpret it you start from the center where it says "F". You then follow the arrows which lead to the left to "P" the product and to the right to "W" the waste. In a cascade like this the waste from a centrifuge is fed back to the previous centrifuge to be recycled. Waste not, want not. You will notice that mass is conserved as we go around the loops. Now I am going to compute the net separative work done by this cascade in two ways. First I am going to add up the work done by each centrifuge taking advantage of the fact that θ does not depend on the stage and is always $1/(1+\alpha)$. Thus to carry out this calculation I will add up the feeds and apply the formula from above. Thus
$$\delta U/L = 3/2(1+2\alpha+2\alpha^2+\alpha^3)(1-\alpha)^2\alpha \sim 9/2(\alpha+1)(1-\alpha)^2.$$
The second way I am going to do the calculation is to suppose that the cascade is a black box which is fed $1+\alpha^3$ and enriches by a factor $\alpha^3$. Thus

$$\delta U/L = 1/2(1+\alpha^3)(1-\alpha^3)/\alpha^3 \sim 9/2(\alpha+1)(1-\alpha)^2$$

This gives the same answer.

I am going to stop this lengthy section here and in the next one I will discuss the implications when it comes to determining how many centrifuges are needed in a cascade to produce a certain number of SWU. I would also like to discuss some more realistic cascades and then tell you what I know about Iran.

V.

In the spring of 1959 I was in residence at the Institute for Advanced Study. One of my fellow residents was Dirac. He was working on a theory that seemed to require what we used to call the mu meson to have spin zero. That it didn't did not seem to concern him. He was also chopping a path in the woods behind the Institute. To where I am not sure. He had an office near mine and near that of Bram Pais. None of the temporary members had phones in their offices. I think it was Oppenheimer's idea that phones might be a distraction. There was a communal phone in the hall that distracted everyone. It rang. It was for Dirac. It was from the New York Times's Walter Sullivan. He had read that Dirac was giving a lecture in New York and wanted an advance copy. Dirac agreed, but then had second thoughts which he wanted to discuss with Pais who happened to be in my office. Pais listened and said that what should Dirac should do is to send the manuscript to Sullivan writing on it, "Do not publish in any form." Dirac stood there silently thinking for several minutes and then asked Pais, "Isn't 'in any form' redundant in that sentence."

Another visitor was a very nice Dutch physicist named Nicolaas Marinus Hugenholtz. My only knock on 'Nico' was his tendency, often at lunch, to make oracular pronouncements with a ponderous Dutch accent. One noon he said, "Wigner is a very remarkable man." "Why?", I asked. "He publishes only one tenth of what he knows," Nico replied. "Nico," I said, " I too am a very remarkable man-I do just the opposite."

Well, as the bishop said to the actress, "Enough, madam, I did not come here to talk."

Below is the diagram of the Whitley toy model.

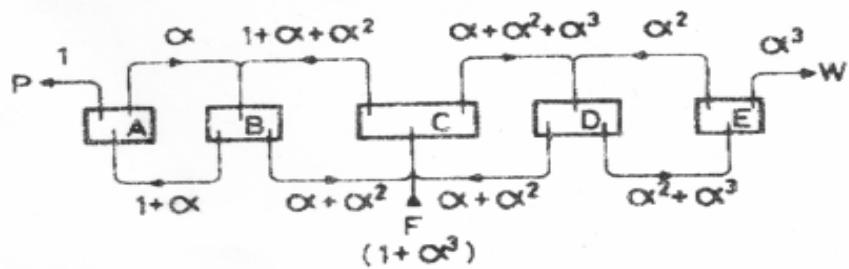

FIG. 16. Simple symmetrical cascade.

Like many good educational toys it still has a lot to teach us.

The first question that struck me is how did it get that way? A cascade like this does not spring full-blown from the forehead of Zeus. You have to turn it on. How you do this? The first step is to plug the centrifuges into the electric power grid. This raises the question of how much power do you use? If I you are making bombs this might not be a worry but if you are using the lightly enriched uranium-LEU-to power electricity producing reactors that is certainly a concern. To answer this I am going to use some numbers that you can find in a letter that Dick Garwin wrote in March/April 2005 *Foreign Affairs*-which is readily available on the web. He notes that a centrifuge's power consumption is about 100 kilowatt hours per SWU. I do not know how much variability there is in this number. He is studying the Iranian situation and makes the assumption that their centrifuges produce 3SWU per year. Later I will discuss this. Here I note that the best modern centrifuges which there the Iranians now have, have separation factors as high as $\alpha \sim 1.5$ and can produce as much as 100 SWU per year. Since there are 8766 hours in a year the power consumption of such an Iranian centrifuge is 34 watts-a dim light bulb. To power an electricity producing reactor for a year you need about 1000 kilograms of Uranium 235. In the last section I noted that the number of SWU required to produce a four percent enriched product was about six. But if we want the U-235 in this product we have to divide by .04 which gives about 150 SWU per kilogram of U-235. If each centrifuge can produce 3 SWU per year then we need 50,000 running for a year. As I mentioned, the modern centrifuges do a lot better, but just to stick with these numbers. We have a power consumption of 50,000x34 watts=1700kilowatts. But such a reactor can put out a million kilowatts so we are, even with these rather primitive centrifuges. well ahead of the game.

Once the centrifuges are started then you can begin to feed in the hex. If you are starting from scratch then the U-235 concentration is about one percent. This is done with product outlet valves closed- " total reflux." The enrichment process begins and one waits until the product stage is at the desired enrichment. Then one begins to extract product by gradually opening the appropriate valves. What is called the "equilibrium time" is the time it takes for the flow of the product reaches half of its final value. This time varies from cascade to cascade but it is the order of hours. Then you are in business.

Before we begin the heavy lifting here is a little bon-bon for you. The modern centrifuges have peripheral speeds of some 500 meters a second. A radius of a cylinder is, say, 10 centimeters. Thus the centrifugal acceleration is about 250,000 that of gravity. Even on Jupiter on which the gravitational acceleration is about 2.5 times greater we still have a big advantage with the centrifuge. In their original paper Lindemann and Aston noted that if you sent a balloon up you could in principle detect the different isotopes of neon. They decided that this was more trouble than it was worth. A brief and superficial scan of the web seems to indicate that the study of the separation of atmospheric isotopes by gravitation is alive and well both here and on Mars.

We can now count the number of stages that lead to enrichment in the toy model. I use the word "stages" rather than "centrifuges" because in real life each stage can, and usually does, consist of several centrifuges in parallel. I will come back to this later. I want to explain how to estimate the number of stages when there are a great many of them and the change from one to the next can be treated as infinitesimal. I am always going to assume and Ideal cascade in which $R_s=\alpha^s R_o$. Note that

$N_s/1-N_s = R_s/(1+R_s)/(1-R_s/(1+R_s)) = R_s$.

This will come in handy shortly. To simplify notation I will call the molecular density of the product at stage s, $N'_s$, and the molecular density of the waste at stage s, $N''_s$, I will also similarly label the R's Taking a clue from the toy model the cascades that I am going to consider have the property that

$$R'_{s-1}=R_s=R''_{s+1}$$

and therefore

$$N'_{s-1}=N_s=N''_{s+1}..$$

Consider

$$N_s'-N_s=\alpha R_s/(1+\alpha R_s)-R_s/(1+R_s)\sim(\alpha-1)N_s(1-N_s)$$

where in the denominator we have used the fact that $\alpha\sim 1$. Like wise

$$N''_s-N_s\sim -(\alpha-1)N_s(1-N_s).$$

We can convert these equations into differential equations. For example

$$dN_s/ds=-(\alpha-1)N_s(1-N_s).$$

This is a quadrature than we can actually perform bmewith the result that the number of enrichment stages required to reach an enrichment $N_s$ or $\alpha^s R_0$

$$=1/(\alpha-1)Ln(N'_s/(1-N'_s)\cdot(1-N_0)/N_0)\sim sLn\alpha/(\alpha-1)\sim s(3-\alpha)/2.$$

If the enrichment is $\alpha^n$, then three is replaced by n.

With the machinery assembled we are in a position to do something useful. We can ask what is the total feed that must flow through the cascade to produce say P kilograms of HEU. This is important to know because it will influence how we design the cascade. We expect that the initial flow must be much larger than P because it contains only about a percent of U-235, while P contains 90 percent or more and furthermore the enrichment factor $\alpha$ is not that far from one. The flow in the cascade varies from stage to stage. Some of the flow goes to waste which is recycled. In the ideal cascade I am going to consider the recycle from stage s goes back into stage s-1 and become part of the feed for stage s. Let us call the flow through stage s, $L_s$. then the total flow is given by

$$\Delta L = \sum_0^S L_s = \int_0^S L(s)ds.$$

It is more convenient to integrate with respect to $N_s$, the molar fraction of U-235. $N_0$ is the one percent and $N_p$ is the 90 percent. Thus

$$\Delta L = \int_{N_0}^{N_P} L(N_s) ds/dN_s \, dN_s.$$

Here $N_s$ is the molar fraction of U-235 in the $s^{th}$ stage. The derivative we have done before.

$$ds/dN_s = 1/(\alpha-1)(N_s(1-N_s)).$$

To find $L(N_s)$ we have to use various facts about the cascade. At each stage the feed is split into a product and a waste with the cut $\theta$ and $1-\theta$ respectively where in the case of $\alpha \sim 1$ we have $\theta \sim 1/(1+\alpha)$. In the cascades we are considering $R_s = \alpha R_{s-1}$, which determines how the N's are related. We can still learn something useful from the toy cascade. Draw a line that cuts though the flux and reflux lines anywhere in the diagram. Then add up the terms you find there-the $\alpha$'s-with a plus sign when you add the terms moving toward the product end and a minus sign when you add up the terms moving towards the waste end. You will find that wherever you make this cut the terms will add up to one. This is not a mathematical trick. It is the conservation of the material that runs though the cascade. Below is one of my homespun drawings of a piece of a more sophisticated cascade but on that is "ideal" in that the same conservation of material applies.

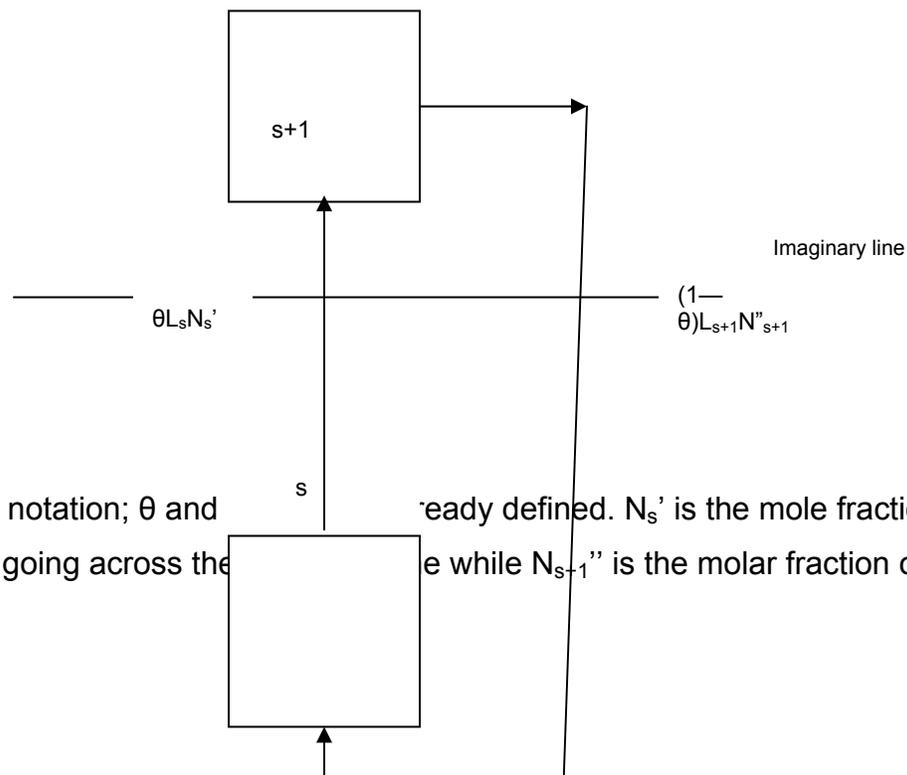

To explain the notation; $\theta$ and ~~~~ready defined. $N_s'$ is the mole fraction of the product going across the ~~~~ while $N_{s+1}''$ is the molar fraction of the waste. By conservation the difference is going to equal $PN_p$ where P is the amount of final product and $N_p$ its molar fraction-say .9 for HEU. So we have the equation

$\theta L_s N'_s - (1-\theta) L_{s+1} N'_s = N_p P$.

I have here used the fact that $N''_{s+1} = N'_s$ which you can verify using the R's. But from the overall conservation we must have

$\theta L_s - (1-\theta) L_{s+1} = P$.

These equations can be combined to get rid of $L_{s+1}$ and leave

$L_s = P/\theta \cdot (N_p - N_s)/(N_s' - N_s)$.

The last parens in the denominator can be written out as

$\alpha R/(1+\alpha R) - R/1+R) = r \cdot (\alpha - 1)/(r+1)(\alpha r+1)$

$\sim (\alpha-1)(N_s(1-N_s)$

where R here is the R appropriate to the $s^{th}$ stage of the enrichment and we have used our approximation $\alpha \sim 1$.

Thus

$L_s = P2/(\alpha-1) \cdot (N_p - N_s)/N_s(1-N_s)$

We are now in a position to evaluate

$$\Delta L = \int_{N_9}^{N_p} L(N_s) ds/dN_s dN_s = P2/(1-\alpha)^2 \int_{N_0}^{N_{Pp}} (N_p - N_s)/(N_s^2 \bullet (1-N_s)^2) dN_s$$

Note that

$$\int (a-x)/(x^2(1-x)^2) dx = (2a-1) Ln(x/(x-1)) + (x-2a+a)/x(x-1)$$

Thus

$\Delta L = 2P/(\alpha-1)^2 [(2N_p -1) \ln(R_p/R_0) + (1-2N_0)/(1-N_0) \cdot (N_p - N_0)/N_0]$

Where as usual

$$R=N/(1-N).$$

The quantity in the square brackets may suggest something to you. Recall that in section II I found the value function by an integration in which I set the additive constants equal to zero, If I don't do this then using the Dirac-Fuchs-Peierls boundary conditions in which the function and its derivatives vanish at $c=c_0$

$$V=(2c-1)Ln((c/1-c))(1-c_c)/c_0))+(c-c_0)(1-2c_0)/c_0(1-c_0)$$

which is just what is in the bracket. In fact this is how Dirac found the value function in the first place. Let me call this value function $V'(N)$. Thus we can write the previous result as

$$\Delta L(P)=2P/(\alpha-1)^2 V'(N_p).$$

I have introduced this notation because we have left out some of the flow. There is also a flow that goes to make the final waste.

$$\Delta L(W)=2W/(\alpha-1)^2 V'(N_w).$$

In general the cascades are operated so that $N_w \sim N_0$. This means that $V'(N_w) \sim 0$. For example if you take $N_w=.0025$, $V'(N_W)=.6$, while with $N_p=.9$, $V'(N_p)=95$. In what follows we shall ignore $V'(N_W)$. We can use what we have learned so far to estimate the number of centrifuges in a cascade designed to do a given task. From the last section if an amount L is fed into an individual centrifuge then $\delta U \sim L(\alpha-1)^2/2\alpha$. Thus

$$\Delta L/\delta U \sim P/L \times 4\alpha/(\alpha-1)^4 V'(N_p).$$

It is however instructive to put a typical $\delta U$ in and see the result. I will take $\delta U=2$ kilogram/year and suppose we want to make 20 kilograms per year of HEU. If we take $\alpha=1.2$ and use the value of $V'(N_p)$ we found before for $N_p=.9$ we find that we need about 24,000 centrifuges in the cascade to do the job. On the other hand if we take $\alpha=1.3$ and $\delta U=3$kg/year, then about 7,000 will do.

    I am going to finish this section with a rather pretty argument that I learned in Whitley. To set the stage I am going to backtrack. In this part I will be following a nice monograph by Douglas Holliday and Milton R.Plesset which dates back to 1966. It is called Elementary Introduction to Isotope Separation

and is the Rand report RM-4938-PR. It is out of print but for $20 Rand will print you your very own copy.

These authors go back to the atmospheric separation of isotopes and its analogy to the centrifuge which we discussed in previously. They write the pressure at a distance from the center of the centrifuge r as

$p(r)=p(0)\exp(M\omega^2 r^2/2kT)$.

Let $M_1$ be the light isotope's mass and $M_2$ that of the heavier. Then

$p_1(r)/p_2(r) = p_1(0)/p_2(0)\exp-(M_2-M_1)\omega^2 r^2/2kT$

The mole fractions are given by

$N=p_1/(p_1+p_2)$ for the lighter isotope and $1-N=p_2/(p_1+p_2)$..

Thus the separation factor α is given by the ratio of the mole fractions at r=0 and r=r which is

$\alpha=\exp((M_2-M_1)\omega^2 r^2/2kT) \sim 1+(M_2-M_1)\omega^2 r^2/2kT$.

But the separative power goes as $(\alpha-1)^2$ which with these approximations goes as $(\omega r)^4 = v^4$ where v is the peripheral speed. If I put in 500m/s for v and 3 proton masses for the mass difference and take T=300 and I come up with α-1~.15.

This is very nice but it may not be relevant to a real centrifuge. What I am now going to tell you I learned from Whitley although the basic idea seems to go back to Dirac. The point is that with the centrifuges that are rotating as fast as these everything gets pushed to the outer surface of the cylinder. You do not have a smoothly varying atmosphere but rather a thin layer at the surface. The geometry effectively becomes planar and the surface has an area of Z2πa where Z is the length of the centrifuge and a its radius. The separation power is proportional to this area. In this configuration the "gravitational" acceleration is equal to

$v^2/a$. Thus α is proportional to $v^2$ instead of $v^4$ and so is the separative power. This appears to apply to real centrifuges, the ones that separate uranium isotopes.

VI.

I am going to begin this section by a recasting much of what has gone before in a simplified kind of executive summary. I want to put things in a bare bones way that will help us to see how to design an actual cascade. There will be some repetition inevitably. If formulae have been proven in previous sections I may just quote the formula and the reference. So here we go.

First I want to treat the single centrifuge. We have seen that the partial pressure ratio is

$$p_1(r)/p_2(r) = p_1(0)/p_2(0) \exp(-(M_2-M_1)\omega^2 r^2/2RT)$$

where all these symbols have been defined previously and if r is the distance to the periphery then $\omega r$ is the peripheral speed $v_p$. If we are dealing with an ideal binary gas then the above can be re-written

$$(N/(1-N))_r = (N/1-N)_0 \exp(-(M_2-M_1)\omega^2 r^2/2RT).$$

The separation factor α is

$$\alpha = (N/(1-N))_r / (N/1-N)_0 = \exp(-(M_2-M_1)\omega^2 r^2/2RT).$$

0r

$$\varepsilon = \alpha - 1 \sim (M_2-M_1)v_p^2/2RT$$

with $R = 8.3 \times 10^7$ ergs/mole K and $M_2 - M_1 \sim 3$ grams/mole. We shall take $T = 300$. The reader may well wonder how does this relate to our previous definition of $\alpha$ where the enrichment was given after a stage as $\alpha R$. But note that

$$(N/(1-N))_r / (N/1-N)_0 = ((\alpha R/(1+\alpha R)/(1- \alpha R/(1+\alpha R))/ R/(1+R)/(1- R/(1+R)) = \alpha$$

The definition in terms of the N's is more general.

      I was describing gas centrifuges to someone who was beginning his study of physics. He asked me a question which I think had been gnawing away at my subconscious. I imagine putting the hex into a cylinder and rotating the cylinder. Why does the gas rotate at the speed of the cylinder? In the things I had read I had not seen this obvious question discussed. It is usually just assumed that it just does. Dick Garwin provided a very nice explanation which I pass on. The solid surface of the cylinder is atomically rough. The outermost gas molecules collide with it and acquire the velocity of the cylinder. This velocity then diffuses inward into the gas the way heat would, and for the same reasons. Hence the gas as a whole acquires the velocity of the cylinder-in effect there is developed a macroscopic viscosity. I am sure that this is discussed somewhere and that I did not pay proper attention to it.

      To evaluate the expression for $\varepsilon$ what should we take for $v_p$? Holliday and Plesset whom I have previously cited make an interesting choice which enables me to say a bit more about the history. Around Christmas of 1938, Otto Hahn and Fritz Strassmann made the first observation of what Lise Meitner and Otto Frisch identified as fission. The neutrons that are emitted with the fission fragments were soon observed and it became immediately clear to many people that a new source of energy was at hand among them was the physical chemist Paul Harteck. He had been born in Vienna in 1902 . At the time of the fission discovery Harteck was a professor in Hamburg. He was also a consultant to German Army Ordnance on explosives. I do not think that he was political, but he, like Willie Sutton, saw where the money was. He had an assistant named

Wilhelm Groth. In April of 1939, Groth and Harteck wrote a letter to German Army Ordnance in which they pointed out the possibility of using nuclear energy as an explosive. This was one of the notices that Army Ordnance had received on this general subject.In June the German physicist Siegfried Flügge had written a widely read popular article in which he had pointed out if all the uranium in a cubic meter of uranium oxide could be fissioned the energy released would be equal to the output of all the power plants in Germany for 11 years. By that fall, the Kaiser Wilhelm Institute for Physics had been taken over to serve as the center of what came to be known as the *Uranverein*-the uranium club. People like Heisenberg and Harteck were drafted to work on the project which in the beginning certainly had as a goal the possibility of making a weapon.

        The project was divided over several institutions. Heisenberg for example was located in Leipzig and wrote two monumental reports on the physics of what became nuclear reactors-reactors which he was never able to build during the war. Harteck and Groth remained in Hamburg and devoted themselves to the problem of separating isotopes. What interests us is what Groth especially was able to do with the centrifuge. In 1943, Harteck invented a novel centrifuge known as a "double centrifuge." These were a pair of centrifuges whose outputs could be connected. They were run at the same rate except that periodically one of them would be run at a slower frequency. Below is a diagram taken from a 1958 report by Groth et al.[14]

---

[14] Groth et al Enrichment of Uranium Isotopes by the Method of Gas Centrifugation, AEC-tr-3412

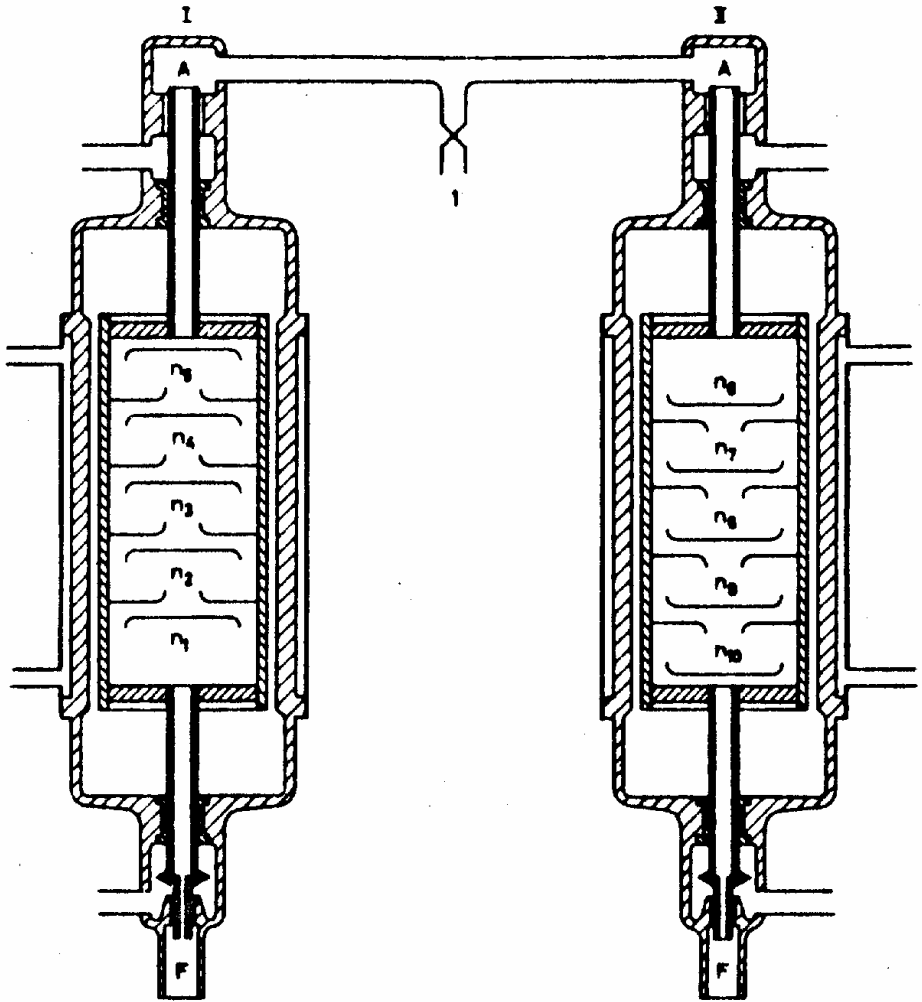

Fig. 1. Action diagram of the rocking procedure

The centrifuges are divided into "chambers." When the centrifuges are run at different speeds the pressure difference produces a current of gas which forces the light isotope out of the second centrifuge and into the first enhancing the enrichment. This is a somewhat primitive version of the counter-current centrifuge design first suggested by Harold Urey in the late 1930's. Urey's design involved a single centrifuge and the currents were maintained thermally,

Mark Walker in his book German National Socialism and the Quest for Nuclear Power[15] writes that the Harteck design doubled the separation capacity. This it did not do but it certainly augmented it. By 1945, Walker tells us, the centrifuges that now Groth was responsible for, were producing 50 grams a day of 15 percent enriched uranium.(This is from a Manhattan Project historical document that Walker found in the National Archives.) Let us translate that into SWU. Using Garwin's table and taking the tails at 0.0025 this enrichment costs 30 SWU per kilogram. Groth's centrifuges allegedly produced 18 kg/year of 15% enriched uranium. This would have required about 540 SWU and hundreds of centrifuges. There was a plant in Celle in Germany where the actual centrifuges were located. I do not know the details of whatever cascade was used but I can tell you the details of the parameters of the centrifuges that were used. The invaluable Whitley in Part II of his review article gives them.[16] It will give us an excellent opportunity to put our formalism to practical use.

Whitley notes that Groth and his collaborators, like the Russian group, had available to them the articles of Beams which they took as their basis. The best of their wartime centrifuges was the so-called UZ111B. Its aluminum alloy tubes had a diameter of 134 millimeters and a length of 635 millimeters and a peripheral speed of 280 meters a second. Thus[17]

$$\varepsilon = \alpha - 1 = 3\,(2.8 \times 10^{4})^2 / 2(8.3 \times 10^7 \times 300) \sim .05$$

Next we want to find the separative work capacity. To ths end we use the formula previously derived

$$\delta U = (L/2) \times (1-\alpha)^2.$$

Here we must be careful. L is often given in terms of kilograms/unit time of hex. For example Holliday and Plesset give for the feed of the UZ111B, 1 gm/minute of hex. But we want the feed of the uranium. The fluorines go along for the ride. Thus we must multiply the L in hex by $238/(238+6 \times 19) \sim 0.68$. If I do this and use the results so far I find that the UZ111B produced about .5 SWU kilograms/year. Holliday and Plesset quote .8. I think they may have forgotten the factor of .68. In the Groth paper cited before there is a figure that shows that the maximum number of SWU which the most advanced centrifuge produced was about 1 kg SWU/year. To produce the 18 kilograms per year of 15% enriched uranium we said requires 540 SWU. Thus if the Celle cascade actually did this it would have required about 1080 centrifuges. Mark Walker has found a letter from Harteck dated May of 1944 which states that if their program received the maximum possible support they could install ten or fifteen double centrifuges at Celle. The numbers don't all add up . It seems evident that the report Walker cites grossly exaggerated the actual situation.

---

[15] Cambridge University Press 1993
[16] Rev.Mod .Phys., Vol.56, No.1, 1984.
[17] In a previous part I pointed out that because the hex gets pushed to the walls of the cylinder the separative capacity may go a $v_p^2$ and not $v_p^4$. The theoretical model used above may not apply and one might actually have to measure the SWU.

Nonethless, is clear that the Germans had made considerable progress on centrifuges under very difficult conditions during the war and this continued for awhile on a research level. I once asked Zippe why these centrifuges had never gone into post war industrial production after around 1958 when Groth got out of the business. Zippe told me that it was a matter of energy. The UZ111B consumed 1.5 kilowatts. This meant that it consumed

1.5 kwx8766 hours/year/.5SWU/kilogram year

~26,000 kw /kg/hour

A modern centrifuge uses something like a thousand times less power. The Russian centrifuges used 2 to 3 watts.

It is interesting to try to understand what made the difference. Whitley tells us that both the Groth and the wartime Beams machines generated a lot of frictional heat. There was a power loss of about 2 kilowatts which created temperature gradients that caused difficulties. The key point was the bearings. They must keep the rotor in an equilibrium position. You don't want it tipping over into the cylinder. At high peripheral velocities the balance is maintained gyroscopically but until these forces take over it must be maintained mechanically. It was just here that the Russian centrifuge was so much better. At the top they used a magnetic bearing which had actually been discussed by

Beams and at the bottom a sort of single needle shaped pivot bearing which was constantly oiled. What is amusing, as Whitley points out, is that both of these bearings had been analyzed at the end of the 19[th] century by a man named S.Evershed who published in 1900 a paper entitled "A frictionless motor meter." In fact Evershed himself had predecessors which inspired a comment in a 1900 meeting by the president of the American Institute of Electrical Engineers "People of old times had very little honesty: they have stolen all our best inventions." Be that as it may, there is a crucial point about all these centrifuges. The material of the cylinders have natural frequencies of vibration. When the rotational frequencies of the rotor matches one of these critical frequencies, serious instabilities can develop. This divides centrifuges into two categories-sub-critical and super-critical. The α's of the supercritical centrifuges are larger but one must deal with the passage though the critical regimes.

In 1956 Max Steenbeck, who had been the leader of the Russian program and its theorist, went to East Germany. He apparently had Communist sympathies all along. Zippe went to West Germany. As I have mentioned previously Zippe, and presumably Steenbeck, were not allowed to take any centrifuge-related documents with them. Steenbeck became a university professor in Jena and as far as I can tell, never worked on centrifuges again. Zippe, on the other hand, had no job to go back to. Very shortly after his return he went to a conference in Amsterdam on centrifuges. It was here that he first realized just how superior the Russian centrifuges were to what was being discussed. One of the leading centrifuge people of the time was the Dutch physicist Joseph Kistemaker. The centrifuges he was working on were horizontal with the bearings at each end. It would be fascinating to know in exactly what way Zippe approached Kistemaker. One does know that after this Zippe was a frequent visitor to Amsterdam and that Kistemaker began developing vertical centrifuges. The Russians had not patented any of this work since they wanted it kept secret. I have read Steenbeck's autobiography and he seems a little unhappy about how this evolved. Many of the ideas were surely his. In any event Zippe began a relationship with the German firm Degussa. Books have been written about Degussa and its unsavory Nazi past. One of its subsidiaries, Auer, supplied the metallic uranium for the Uranverein. Some dangerous steps in the process were performed by concentration camp inmates. Heisenberg surely knew this. Another of the Degussa companies made the poison gas Zyklon B that was used in the extermination camps. It would seem that by 1958 the company had cleaned up its act enough so that, still having a relationship with Degussa, Zippe spent two years in Virginia with Beams. The details of the machine he developed there are known because he published them in 1960. It had a peripheral speed of 350 meters a second, an $\alpha$ of 1.09 and a separative work power of .39 kgSWU/year and a power loss of only 10 watts. This is the last modern centrifuge about which the details are known. The curtain then came down.

We are now beginning the approach that will land us in Iran, which is what motivated all of this in the first place. We have seen that in 1974, A.Q.Khan returned to Pakistan with the plans for the latest centrifuge in addition to come component parts. The details of this centrifuge-the P1-such as they are known, are classified. They must be similar to the details of the centrifuge that Khan trafficked to Libya and these are presumably known since the Libyans turned them over to us. This centrifuge is very probably a variant of the SNOR/CNOR centrifuge that was designed in Holland. Kistemaker was the principle here, and I think Zippe consulted from time to time. The figures I have seen for this centrifuge is a peripheral speed of 350 meters/sec and a SWU that varies- depending on which source you read, of between 1 and 5. The low end is more likely. The Iranians have confirmed the 350 meters/second and a frequency of 64,000rpm[18]. If we use the "barometric" forumula given above then it would indicate that these centrifuges have an $\alpha$ of 1.07.As of this writing-(November 2007) –the Iranians claim to have fed 1240 kg of hex into their centrifuge complex. This amounts to 1240x238/(238+6X19)kg= 838 kg of uranium. They claim to have enriched this to 3.5%. How much enriched uranium came out of the cascade assuming conservation at every step?

---

[18] See the fascinating interview dated April 12,2006 of Gholamreza Aqazadeh, the head of Iran's Atomic Energy Organization-www.armscontrolwonk.com/file_download/32

From conservation we have the relation

$$P=F(N_0-N_w)/(N_p-N_w)$$

where we take F to be the feed of uranium 843kg. We shall need a choice for $N_w$. As a rule this, which is at the behest of the cascade designer, is about .002. But it would appear that the Iranians actually chose-at least as of 2006-the rather large number 0.004. Why I do not know. I will take 0.004.[19] Thus we have

$$P= 843\times(.007-.004)/(.035-.004)kg=82kg$$

This is the total amount of uranium. To find the amount of uranium 235 we have to multiply by .035 which gives 2.9 kilograms. We start with a feed of 843 kilograms of uranium and end up with a total of 81 kg of uranium going through the last product centrifuge. Where does the rest go? Note that

$$W/P=(N_p-N_0)/(N_0-N_w).$$

With the numbers assumed this is 9.3 so there are in this set up 9.3x82kg=763kg of waste. Note that (9.3x81+81)kg=843kg it adds up.

Actually, the Iranians had as of 2007 eighteen cascades of 164 centrifuges each, connected in parallel. Each of these cascades contributes to a common product scheme. To analyze the individual cascades we must divide the above numbers by 18;ie, a feed of 47kg per cascade divided into a product of 6kg and a waste of 41kg.

There is a very important lesson here for cascade design. We feed into the first stage a lot more uranium than goes through the product stage or the waste stage for that matter. Why is this important?

For reasons of economy we want the cascade to consist of essentially identical centrifuges. We want to produce them like clones. But if all the stages are individual centrifuges in series then the flow though the sth centrifuge in the product series is given by

$$L_s=P2/(\alpha-1)\cdot(N_p-N_s)/N_s(1-N_s)$$

---

[19] I am grateful to Dick Garwin for communications about this.

See the previous screed for the derivation. Here the s's are positive integers that run from zero to the final S which produces the last enrichment. Clearly $L_s$ is tending towards zero as $N_s$ approaches $N_p$. Indeed, as $N_s$ is an increasing function of s, $L_s$ is a continually decreasing function. But we want our centrifuges to be all alike and to deal with the same $L_s$ throughout the cascade. How can we possibly arrange this?  Below are three choices for the very simplistic "cascade" of three centrifuges. In all three cases I will assume that the "cut" is 1/2;ie, waste and product are equally divided at each stage. I will not recycle the waste. The first is a connection in series. Each centrifuge can handle a stream L, or less. The enrichment factor is as usual α.

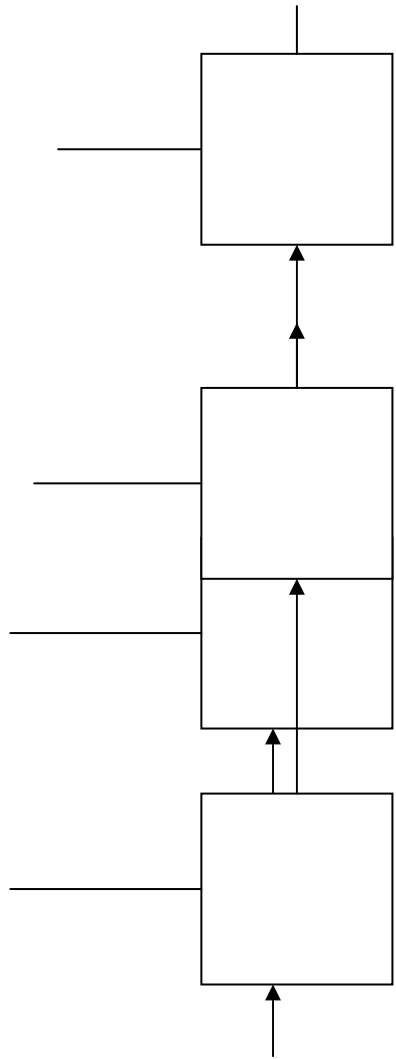

The lower centrifuge is fed a stream L which is divided in two. The final product stream is L/4 but it is enriched by a factor of $\alpha^3$. Next we show three centrifuges in parallel.

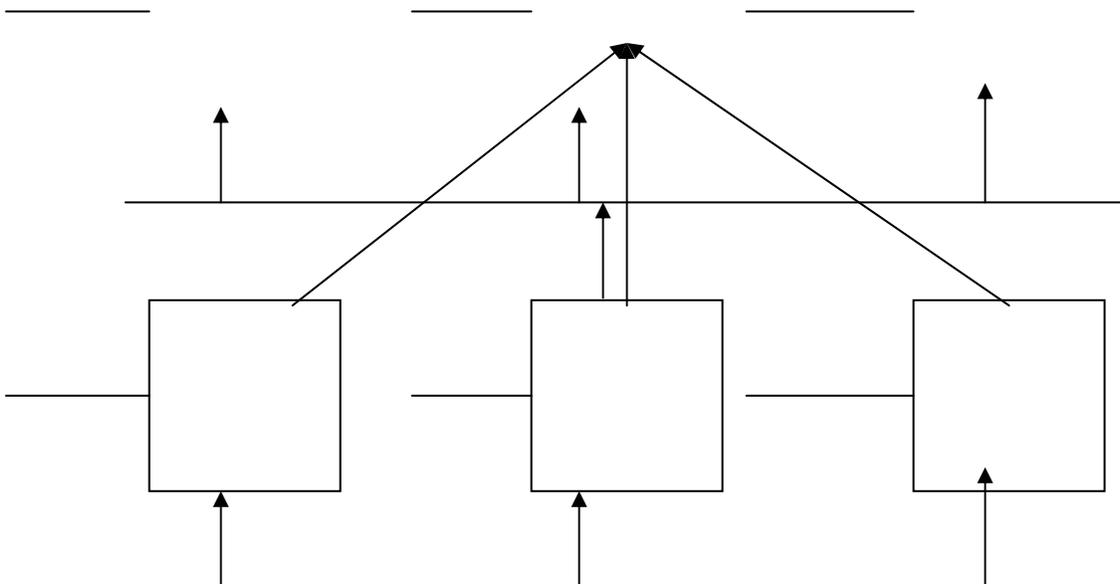

With this setup we can feed in 3L which is divided three ways. Each centrifuge creates L/2 of product for a total of 3/2L, but only enriched by $\alpha$.

Finally we have a mixed series and parallel cascade.

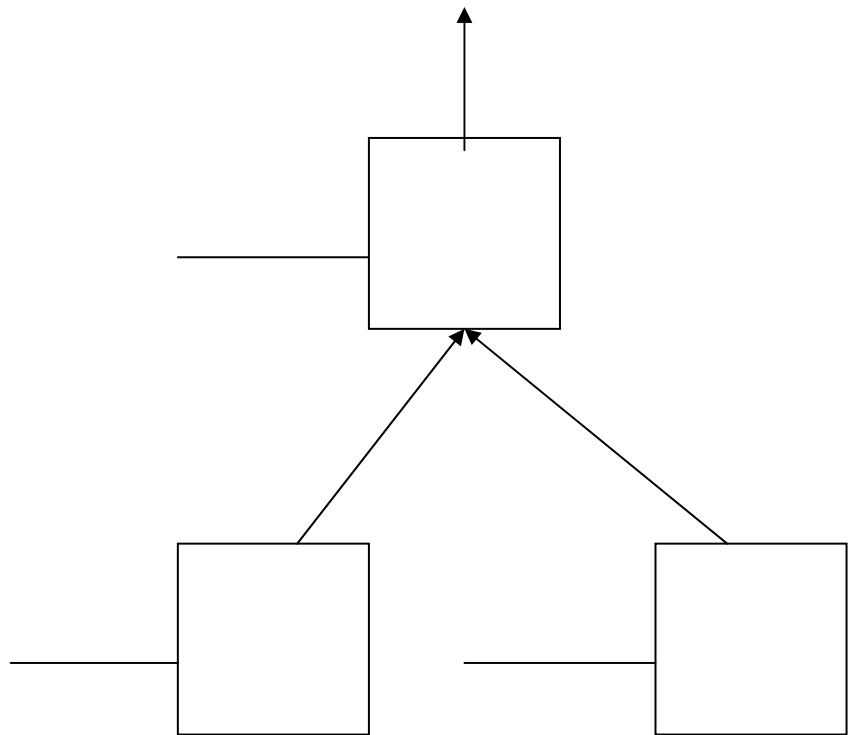

We can feed in 2L, with L going to each of the lower centrifuges. L goes into the final centrifuge and L/2 product comes out enriched by $\alpha^2$. It is this mixed configuration one finds in general. Below is a typical cascade.

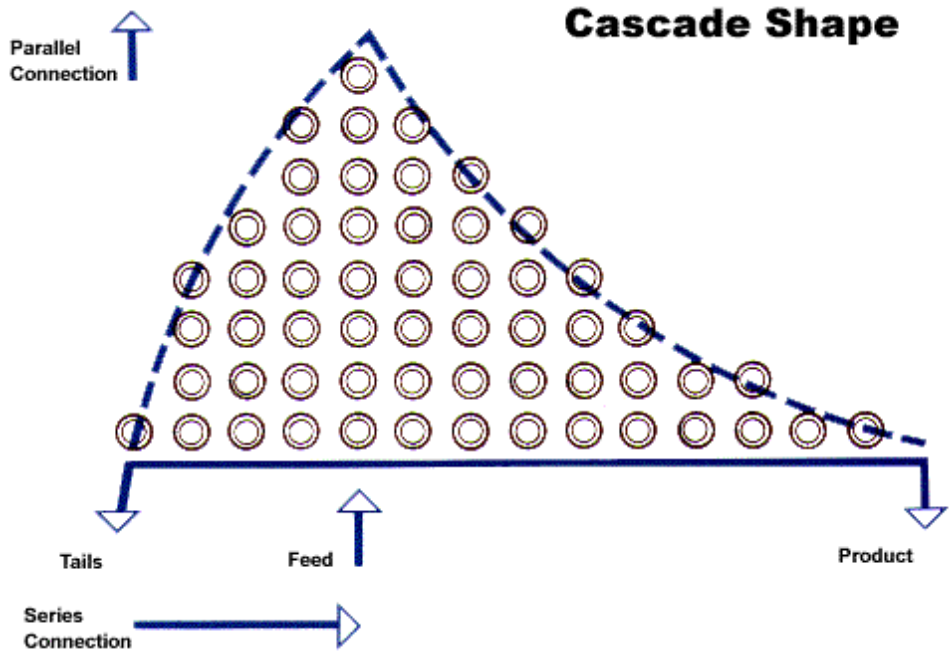

In this and the one below the waste or tails centrifuges are shown. The convention is that the product stages are numbered from zero to S, while the tails stages are numbered from -1 down to what ever the final number of tail stages is. Below is another diagram showing both the product and tails configurations.

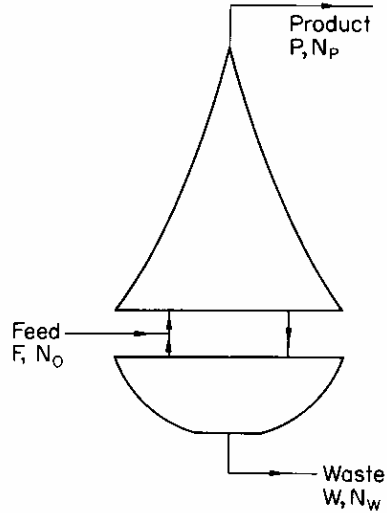

What I want to do now is to collect all the relevant formulae for analyzing the cascades assumed to be ideal in one place. We have either given the derivations or the steps required to fill them in can be read off from what we have done. I will give them for α not necessarily ~1. The transition to the infinitesimal case is trivial. . The flow for the product and waste individually involves the value functions. I will not give them [20] in detail. But the total flow with both waste and product simplifies making use of the conservation of uranium. It is given by

$$L_{tot}=(\alpha+1)/((\alpha-1)\ln(\alpha))(P(2N_p-1)Ln(R_p/R_0)+W(2N_w-1)\ln(R_w/R_0))$$

I will give the formulae for the number of stages in these cascades-note it is stages and not centrifuges. Thus

$N(product)=Ln(R_p/R_0)/\ln(\alpha)$

$N(waste)=(Ln(R_0/R_w)/Ln(\alpha))-1$

$N(total)=(Ln(R_p/R_w)/\ln(\alpha))-1$ .

Using conservation we have

$F=P(N_p-N_w)/(N_0-N_w)$

---

[20] For example with $U(P)=P((2N_p-1)Ln(R_p/R_0)+(N_p-N_0)(1-2N_0)/N_0(1-N_0))$ we have for the product flow $((\alpha+1)/(\alpha-1)\ln(\alpha))U(P)$.

and

$W = P(N_p - N_0)/(N_0 - N_w)$.

Finally we want the number of centrifuges in the product and tail stages. Define U(N)

$U(N) = ((2N-1)Ln(R_N/R_0) + (N-N_0)(1-2N_0)/N_0(1-N_0))$

Then the number of product centrifuges is given by

$PU(N_p)/SWU$ kg/yr/centrifuge, while the number of waste centrifuges is given by $WU(N_w)/SWU$ kg/yr/centrifuge.

As an exercise let us build a cascade that will separate 1kg/day of 4% enriched uranium. I will assume that $N_w$=.002. Using the last two formulae, F=7.6kg/day and W=6.6 kg/day. To find the number of centrifuges I need to make an assumption about the SWU. I will assume 1kg/yr/centrifuge. You can change this to your favorite number as you see fit. To find the number of centrifuges we will use the numbers in this case U(.04)=3 and U(.002)=.56. If I put the numbers in I find that we need 1095 centrifuges in the product stage and 203 in the waste stage.

In the fascinating interview with the Iranian director of the program he tells us enough about the 164 centrifuge Iranian cascade so that Dick Garwin was able to analyze it . He finds that α=1.215 and that the number of produce stages is 7.26 while the number of waste or stripping stages is 5.91. I do not understand the discrepancy between the α found this way and the one using the barometric formula. This way sticks close to the empirical facts.

Dick also estimated the SWU/year of the centrifuges. The analysis goes as follows. The 164 centrifuge cascade produces, it is said, 7g/hr of 3.5% enriched product or .245g/hr of U-235 or 2.147 kg/yr Assuming a 0.36% waste the SWU/kg of U-235, by the magic table, is 111 kg/yr. Thus this cascade produces 111x2.147SWU/yr=238 SWU/yr. If you divide this by 164 you get 1.45 SWU/yr per centrifuge. This is a useful number to know when one discusses the potential of the Iranian program. I have read numbers that vary from 1 to 5 SWU/yr.

VII

Here is a Russian riddle. Which is more important the Sun or the Moon? The Moon, evidently, because the Sun shines during the day when it is light anyway. There is nothing like a good riddle to produce deep thought, or thought anyway. Here is a riddle that bothered me. I thank Dick Garwin for providing the answer. I am going, as is my wont, to give a more verbose version of what he told me. Anyway, here is the riddle.

Many commentators on the Iranian scene put the matter as follows. They say that if the Iranians want to go from making LEU-lightly enriched uranium-to HEU-heavily enriched uranium-what they need to do is to keep the hex in the centrifuge longer. The image that comes to mind is that the centrifugal force acting differentially on the two isotopes will keep pushing the heavier isotope further and further from the axis as compared to the light isotope and hence the separation gets better and better. I suspect this works when you separate the cream from the milk. On the other hand those of you who have been following my several screeds know that gas centrifuges don't work like this. To each centrifuge there is a separation factor $\alpha$, which depends on properties of the centrifuge such as the peripheral velocity. The centrifuge enriches the mixture R by the factor $\alpha R$ and deriches it-if there is such a word-by the factor $R/\alpha$, where $\alpha \sim 1$. This is all the enrichment you are going to get from a single gas centrifuge no matter how long it operates. You need a cascade. The riddle is what is wrong with the first picture?

To see what is going on I take you back to near the beginning when I spoke about separating isotopes of say neon in the atmosphere by gravitation. I want to review that here. To find how pressure varies as a function of the height z above the atmosphere what you do is to imagine a little rectangular box in the gas with surfaces of area A. One then makes an analysis of the forces. There is a force from below due to the hydrostatic pressure. But there are two forces from above, one from the pressure and the other from gravity weighing down from above. Let us call the density of the gas $\rho$, Then the pressure difference is determined by the equation

$$dp=-\rho g dz,$$

where g is the gravitational acceleration which we replace by $v^2/r$ when we consider the centrifuge. This formula applies liquids as well as gasses. But now we make the next step for the gas by assuming that thermal equilibrium has been reached at a temperature T. Thus

$$\rho=Mp/RT.$$

Now we can integrate to find

$$p=p_0 \exp(-Mgh/RT)$$

where h is the height above sea-level. For the centrifuge you integrate

$$d\rho/\rho=-Mv^2/RT dr/r=-M\omega^2/RT r dr$$

where r is the distance from the edge of the centrifuge and $\omega$ is the angular frequency. Thus for as given species the density falls off from the edge-"sea level" to the center by

$$\rho=\rho_0 \exp(-Mv_0^2/2RT)$$

where $v_o$ is the peripheral velocity. The moral is that all of the species never ends up at the edge no matter how long you centrifuge because the gas pressure keeps it away.

Mahdi Obeidi is an Iraqi , now living in a undisclosed location in America. He had done his training at the Colorado School of Mines in Golden. He returned to Iraq to take up a career in petroleum engineering. In 1975, it was suggested that he would have a more interesting career if we switched to nuclear engineering, which he did. He began by working on a Russian supplied reactor that was having all kinds of problems. He was sent to Italy for further training and at least one contact he made there had important later implications. In 1979 Saddam took over the country in a bloody coup and soon the French began delivering their Osiris reactor to Iran. It was made clear to

Obeidi that this reactor was going to be used to make material suitable for nuclear weapons. The reactor was destroyed by the Israelis in 1981 but enough highly enriched and nearly highly enriched uranium had survived that attack to make at least one nuclear device. This required enriching the uranium and in 1984 Obeidi began working on a program already started to make barriers to be used in gaseous diffusion enrichment. Ultimately his group made a very good barrier which was never used. One of the reasons was that Obeidi had turned his attention to centrifuges.

By this time Iraq was deeply engaged in the Iranian war and Saddam had become fixated on making a nuclear weapon. In his book A Bomb in My Garden,[21] Obeidi describes a meeting later with Saddam which makes it clear that Saddam did not have any clue as to what this involved. But he had put his son-in-law Hussein Kamal in charge. I was reminded of Stalin's putting Lavrenti Beria in charge of the Soviet program. He and Kamal met similar fates. Beria was executed in 1953 after Stalin's death and Kamel was executed by Saddam's henchmen in 1996 when he took at face value Saddam's invitation to return to Iraq after escaping with his family to Jordan.

Readers now know of my interest in Gernot Zippe. You will have learned that Zippe as a Russian prisoner of war was ordered with some other prisoners to separate uranium isotopes. You will have learned that Zippe had never had any prior contact with centrifuges but that he and the physicist Max Steenbeck, who also had no prior knowledge of centrifuges, created what became the prototype for all the gaseous centrifuges that are now in use. You will have learned that when Zippe was released from the Russia in 1956 he began disseminating this information. You will also have learned that he soon after spent a couple of years at the University of Virginia working with Jesse Beams. He and beams wrote a report describing this work and somehow Obeidi heard of it. He was then sent to the United States to get a hold of it. He went to the University of Virginia and was able to persuade a librarian to have a brief look at it. The most important thing that he learned was that a copy was in Italy in the hands of someone he had known there. He went to Italy and arranged for a copy to be sent to Iraq.

---

[21] Wiley, New York, 2005

This was a web that Obeidi describes of suppliers to Iraq of vital plans and parts for making a centrifuge. As far as I can tell from his account A.Q.Khan had nothing to do with it, but there were plenty of willing suppliers only too happy to accept Iranian money-a few experts even coming to Iran to help out. The net result of all of this is that by 1990, Obeidi and his people had constructed a prototype centrifuge that was a least comparable to what was available elsewhere. To give some idea, it was capable of about thirteen hundred revolutions a second which is about what Zippe's Russian centrifuges produced. The Iraqis were about to make a cascade out of these when the first Gulf War put a stop to it. Obeidi describes listening to Colin Powell at the UN in 2003. He knew that the aluminum tubes Powell was talking about had the wrong dimensions for centrifuges, and besides, the Iraqia had abandoned aluminum for carbon fiber tubes-much superior- a dozen years earlier. In 1992, when the weapons inspectors came to Iraq, Obeidi buried all the plans and some prototype parts in his garden. After the invasion he turned them over to the Americans.

I am not is a position to know how much of Obeidi's story is literally true. But I would like to call your attention to one technical point which now seems to have some relevance when it comes to Iran. This has to with what are in the centrifuge business referred to as "bellows." I do not know who coined this term of art but as far as I can see, the first person to use them in centrifuge design was Zippe. In the 1940's Dirac produced his formula for the maximum output that a given centrifuge could produce. In particular the amount is proportional to the surface area of the rotor-the cylinder. This means that for a given type of rotor the amount is proportional to its length. With Dirac's assumptions it is also proportional to the fourth power of the peripheral speed. I do not think that his assumptions apply to the actual case, something I have discussed in one of my previous sections.

To recapitulate that discussion. At some speed-and I do not know in practice what speed this is-essentially all the hex is driven to the wall of the rotor. The geometry becomes essentially planar so if the radius of the rotor is r and its length d the area mentioned above is sensibly 2πrd. But suppose want our centrifuge to produce a certain centrifugal force characterized by some specified acceleration g. Then $r=v^2/g$ so that the output goes as $v^2$ and not $v^4$. In any event for a fixed v, the longer the centrifuge the better.

In practice one makes long centrifuges by putting together shorter ones. Enter the bellows. The dread of all centrifuge people is that the rotor starts to wobble. At these speeds things get rapidly out of control and you have an explosion. Zippe's idea was to join two shorter rotors together with a metallic cylinder that had some flexibility so that it could dampen potential wobbles. These cylinder's are what are known as bellows. There are generally made of maraging steel-steel that contains metallic alloys and not carbon. They are fiendishly difficult to design. The Iraqis got them from a German supplier. I bring this up because of news from Iran. The claim is that they are now using carbon fibers for their rotors. This has enabled them to greatly increase the peripheral speed. By what I have been saying, this enables them to shorten the rotor apparently to the extent that they can do without the bellows. Their centrifuges which, at least the newer ones, are made using carbon fiber have peripheral speeds of 600 meters a second. How many of the new centrifuges they have in their cascades I do not know, but they claim to have enriched about a metric ton of 3.5% enriched uranium. No amount of uranium with this enrichment can be used to make a nuclear weapon. You must enrich to about 90%. There are two ways of doing this. You can reconfigure the cascade to take natural uranium and enrich it to 90% or you can start with a batch of 3.5% enriched uranium and reconfigure the cascade to complete the job, Either one would require work that could easily be observed by an inspector if any were allowed on the premises, which at present they are not. It would take probably some months to do such a reconfiguration and then once it is done maybe a year to produce highly enriched uranium. It is also possible that the Iranians have a secret facility-centrifuge cascades are relatively easy to hide-which is already producing highly enriched uranium. Nothing coming from Iran shows any signs of conciliation when it comes to nuclear activities. These are dangerous times.